\documentclass[aps,preprint]{revtex4}%
\usepackage{amssymb}
\usepackage{amsfonts}
\usepackage{amsmath}
\usepackage{graphicx}%
\providecommand{\U}[1]{\protect\rule{.1in}{.1in}}

\begin{document}
\title{Extended gauge theory and gauged free differential algebras}
\author{P. Salgado$^{1,2,}$}
\email{pasalgad@udec.cl}
\author{S. Salgado$^{1,2,3,}$}
\email{sesalgado@udec.cl}
\affiliation{$^{1}$Departamento de F\'{\i}sica, Universidad de Concepci\'{o}n, Casilla
160-C, Concepci\'{o}n, Chile}
\affiliation{$^{2}$Departamento de F\'{\i}sica Te\'{o}rica, Universidad de Valladolid,
Paseo de Bel\'{e}n 7, 47011 Valladolid, Spain}
\affiliation{$^{3}$Max-Planck-Institut f\"{u}r Physik, F\"{o}hringer Ring 6, 80805
M\"{u}nchen, Germany}

\begin{abstract}
Recently, Antoniadis, Konitopoulos and Savvidy introduced, in the context of
the so-called extended gauge theory, a procedure to construct background-free
gauge invariants, using non-abelian gauge potentials described by higher
degree forms.

In this article it is shown that the extended invariants found by Antoniadis,
Konitopoulos and Savvidy can be constructed from an algebraic structure known
as free differential algebra. In other words, we show that the above mentioned
non abelian gauge theory, where the gauge fields are described by $p$-forms
with $p\geq2,$ can be obtained by gauging free differential algebras.

\end{abstract}
\maketitle

\section{Introduction}

Higher gauge theory \cite{epjs,sav4,sav5,savv8,savv1,savv2,savv3,ims} is an
extension of ordinary gauge theory, where the gauge potentials and their gauge
curvatures are higher degree forms. It is believed that higher gauge theories
describe the dynamics of higher dimensional extended objects thought to be the
basic building blocks of fundamental interactions. \ 

The basic field of the abelian higher gauge theory, originated in
supergravity, is a $p$-form gauge potential $A$ whose $\left(  p+1\right)
$-form curvature is given by $F=\mathrm{d}A$ from which the Lagrangian and the
action of the theory can be constructed. This abelian theory is known in the
specialized literature as $p$-form electrodynamics and it is endowed with a
local gauge symmetry with the transformation law $A\rightarrow A^{\prime
}=A+\mathrm{d}\varphi$ for some $\left(  p-1\right)  $-form $\varphi$.

The natural question is: does there exist a non-abelian higher gauge theory?
To answer this question it is interesting to remember that the points of a
curve have a natural order and the definition of the parallel transport along
a given curve indeed makes use of this order. However, for higher dimensional
submanifolds such a canonical order is not available. This lack of natural
order led to C. Teitelboim in Ref. \cite{teit} to the formulation of a no-go
theorem, ruling out the existence of non-abelian gauge theories for extended objects.

Recent attempts to circumvent this theorem has been carried out in Refs.
\cite{epjs,sav4,sav5,savv8,savv1,savv2,savv3,ims}. In particular, in
Refs.~\cite{epjs,sav4,sav5,savv8}, were found invariants similar to the
Pontryagin--Chern forms $\mathcal{P}_{2n}$ in non-abelian tensor gauge field
theory, denoted by $\Gamma_{2n+p}$ with $p=3,4,6,8$. \ Since $\mathrm{d}%
\Gamma_{2n+p}=0,$ we can write $\Gamma_{2n+p}=\mathrm{d}\mathfrak{C}%
_{\mathrm{ChSAS}}^{\left(  2n+p-1\right)  }$. In the same references were
found explicit expressions for these invariants in terms of higher order
polynomials of the curvature forms. As with standard Chern--Simons forms, the
secondary forms $\mathfrak{C}_{\mathrm{ChSAS}}^{\left(  2n+p-1\right)  }$ are
background-free, quasi-invariant and only locally defined (and therefore
defined only up to boundary terms, $\mathfrak{C}_{\mathrm{ChSAS}}^{\left(
2n+p-1\right)  }\sim\mathfrak{C}_{\mathrm{ChSAS}}^{\left(  2n+p-1\right)
}+\mathrm{d}\sigma^{\left(  2n+p-2\right)  }$).

The purpose of this paper is to show that the invariants introduced in
Refs.~\cite{epjs,sav4,sav5,savv8} can be constructed from a gauged free
differential algebra.

This paper is organized as follows: In Section~2, we briefly review the
extended gauge theory developed in Refs.
\cite{epjs,sav4,sav5,savv8,savv1,savv2,savv3,ims}. In Section~3, we will make
a short review about free differential algebras and their gauging. Section 4
contains the results of the main objective of this work, namely: to show that
the algebraic structure known as free differential algebras (FDA), allows to
formulate a theory of non-abelian gauge with gauge fields described by
$p$-forms with $p\geq2$ and to prove that the extended invariants found in
Refs. \cite{epjs,sav4,sav5,savv8} can be constructed by gauging free
differential algebras. We finish in Section~5 with some final remarks and
considerations on future possible developments.

\section{Chern--Simons--Antoniadis--Savvidy (ChSAS) forms}

In this section we briefly review the extended gauge theory developed in Refs.
\cite{epjs,sav4,sav5,savv8}.

\subsection{Chern--Simons forms}

The Pontryagin--Chern forms $\mathcal{P}_{2n+2}=\left\langle F^{n+1}%
\right\rangle $ satisfy the condition $\mathrm{d}\mathcal{P}_{2n+2}=0$, where
$F=\mathrm{d}A+A^{2}$ is the $2$-form field strength of the $1$-form gauge
field $A$. From the Poincar\'{e} lemma, we know that locally there exists a
$\left(  2n+1\right)  $-form $\mathcal{C}_{2n+1}$ such that $\mathcal{P}%
_{2n+2}=\mathrm{d}\mathcal{C}_{2n+1}$. This $\left(  2n+1\right)  $-form
$\mathcal{C}_{2n+1}$ is called a Chern--Simons form which is quasi-invariant
under gauge transformations~\cite{Nakahara}.

Using the Chern--Weil theorem we can find an explicit expression for the
Chern--Simons forms. In fact: let $A^{(0)}$\textbf{\ }and\textbf{ \ }$A^{(1)}$
be two one-form gauge connections on a fiber bundle over a $\left(
2n+1\right)  $-dimensional base manifold $M$, and let $F^{(0)}$\textbf{\ }%
and\textbf{ }$F^{(1)}$ be the corresponding curvatures. Then, the difference
of Pontryagin--Chern forms is exact%
\begin{equation}
\left\langle \left[  F^{(1)}\right]  ^{n+1}\right\rangle -\left\langle \left[
F^{(0)}\right]  ^{n+1}\right\rangle =\mathrm{d}\mathcal{T}^{\left(
2n+1\right)  }\left(  A^{(1)},A^{(0)}\right)  , \label{cwt}%
\end{equation}
where%
\begin{equation}
\mathcal{T}^{\left(  2n+1\right)  }\left(  A^{(1)},A^{(0)}\right)  =\left(
n+1\right)  \int_{0}^{1}\mathrm{d}t\left\langle \Theta F_{t}^{n}\right\rangle
, \label{tra'}%
\end{equation}
is called a transgression $\left(  2n+1\right)  $-form, with $\Theta
=A^{(1)}-A^{(0)}$ and \ $A_{t}=A^{(0)}+t$ $\Theta$. The $2$-form $F_{t}$
stands for the field-strength of the $1$-form connection $A_{t}$,
$F_{t}=\mathrm{d}A_{t}+A_{t}A_{t}$.\ Setting $A^{(0)}=0$\textbf{\ }and\textbf{
}$A^{(1)}=A$ in (\ref{tra'}), we obtain the well known Chern--Simons $\left(
2n+1\right)  $-form%
\begin{equation}
\mathcal{C}_{2n+1}\left(  A\right)  =\mathcal{T}^{\left(  2n+1\right)
}\left(  A,0\right)  =(n+1)\int_{0}^{1}\mathrm{d}t\langle A\left(
t\mathrm{d}A+t^{2}A^{2}\right)  ^{n}\rangle. \label{chs}%
\end{equation}
From the Chern--Weil theorem it is straightforward to show that under gauge
transformations the Chern--Simons forms are quasi-invariant. However, it is
important to stress that since a connection cannot be globally set to zero
unless the bundle (topology) is trivial, Chern--Simons forms turn out to be
only locally defined.

\subsection{Non-abelian tensor gauge fields}

The idea of extending the Yang--Mills fields to higher rank tensor gauge
fields was used in Ref. \cite{epjs,sav4,sav5,savv8} to construct gauge
invariant and metric independent forms in higher dimensions. These forms are
analogous to the Pontryagin--Chern forms in Yang--Mills gauge theory.

\subsubsection{ChSAS forms in $(2n+2)$-dimensions}

The first series of exact $(2n+3)$-forms is given by%
\begin{equation}
\Gamma_{2n+3}=\langle F^{n},F_{3}\rangle=\mathrm{d}\mathfrak{C}%
_{\mathrm{ChSAS}}^{\left(  2n+2\right)  },
\end{equation}
where $F_{3}=\mathrm{d}A_{2}+[A,A_{2}]$ is the $3$-form field-strength tensor
for the $2$-rank gauge field $A_{2}=\frac{1}{2}B_{\mu\nu}\otimes
\mathrm{d}x^{\mu}\wedge\mathrm{d}x^{\nu}=\frac{1}{2}B^{a}{}_{\mu\nu}%
T_{a}\otimes\mathrm{d}x^{\mu}\wedge\mathrm{d}x^{\nu}$ and satisfy the Bianchi
identities, $\mathrm{D}F_{3}+[A_{2},F]=0.$ Under gauge transformations, the
gauge potential $A_{2}$ and the corresponding curvature transform
as~\cite{epjs}%
\begin{align}
\delta A_{2}  &  =\mathrm{D}\xi_{1}+[A_{2},\xi_{0}],\\
\delta F_{3}  &  =\mathrm{D}(\delta A_{2})+\left[  \delta A,A_{2}\right]  ,
\end{align}
where $\xi_{0}=\xi^{a}T_{a}$ is a $0$-form gauge parameter and $\xi_{1}%
=\xi^{a}{}_{\mu}T_{a}\otimes\mathrm{d}x^{\mu}$ is a $1$-form gauge parameter.

Using the Chern--Weil theorem, we can find an explicit expression for the
Chern--Simons form. In fact: Let $A^{(0)}$ and $A^{(1)}$ be two gauge
connection 1-forms, and let\textbf{\ }$F^{(0)}$\textbf{\ }and\textbf{
}$F^{(1)}$ be their corresponding curvature $2$-forms. \ Let $A_{2}^{(0)}$ and
$A_{2}^{(1)}$ be two gauge connection $2$-forms and let $F_{3}^{(0)}$ and
$F_{3}^{(1)}$ be their corresponding curvature $3$-forms. Then, the difference
$\Gamma_{2n+3}^{(1)}-\Gamma_{2n+3}^{(0)}$ is an exact form%
\begin{equation}
\Gamma_{2n+3}^{(1)}-\Gamma_{2n+3}^{(0)}=\langle\left[  F^{(1)}\right]
^{n}F_{3}^{(1)}\rangle-\langle\left[  F^{(0)}\right]  ^{n}F_{3}^{(0)}%
\rangle=\mathrm{d}\mathfrak{T}^{\left(  2n+2\right)  }(A^{(0)},A_{2}%
^{(0)};A^{(1)},A_{2}^{(1)}), \label{sixBBB}%
\end{equation}
where%
\begin{equation}
\mathfrak{T}^{\left(  2n+2\right)  }(A^{(0)},A_{2}^{(0)};A^{(1)},A_{2}%
^{(1)})=\int_{0}^{1}\mathrm{d}t\left(  n\langle F^{n-1},\Theta,F_{3t}%
\rangle+\langle F_{t}^{n},\Phi\rangle\right)  , \label{sieben}%
\end{equation}
with $\Phi=A_{2}^{(1)}-A_{2}^{(0)}$, is what we call Antoniadis--Savvidy (AS)
transgression form.

Using the procedure followed in the case of Chern--Simons forms, we define the
$\left(  2n+2\right)  $-ChSAS form as%
\begin{align}
\mathfrak{C}_{\mathrm{ChSAS}}^{\left(  2n+2\right)  }  &  =\mathfrak{T}%
^{\left(  2n+2\right)  }(A,A_{2};0,0)=\int_{0}^{1}\mathrm{d}t\langle
nAF_{t}^{n-1}F_{3t}+A_{2}F_{t}^{n}\rangle\nonumber\\
&  =\left\langle F^{n},A_{2}\right\rangle +\mathrm{d}\varphi_{2n+1}.
\label{acht}%
\end{align}
This result is analogous to the usual Chern--Simons form~\textbf{(}%
\ref{chs}\textbf{)}, but in even dimensions~\cite{ims}. It is interesting to
notice that transgression forms (both, standard ones and the above
generalization) are defined globally on the spacetime basis manifold of the
principal bundle, and are off-shell gauge invariant. Chern--Simons forms
(both, standard ones and the AS generalization) are locally defined and are
off-shell gauge invariant only up to boundary terms (i.e., quasi-invariants).

\subsubsection{ChSAS forms in $(2n+3)$-dimensions}

The second series of invariant forms is defined in $2n+4$ dimensions and is
given by%
\begin{equation}
\Gamma_{2n+4}=\langle F^{n},F_{4}\rangle=\mathrm{d}\mathfrak{C}%
_{\mathrm{ChSAS}}^{\left(  2n+3\right)  }, \label{neun}%
\end{equation}
where the corresponding $(2n+3)$-form $\mathfrak{C}_{\mathrm{ChSAS}}^{\left(
2n+3\right)  }$ is defined in terms of the $4$-form $F_{4}=\mathrm{d}%
A_{3}+[A,A_{3}]$ field-strength tensor for the rank-$3$ gauge field $A_{3}$.
In fact,\ following the procedure shown in the above subsection, we define the
$\left(  2n+3\right)  $-ChSAS form as%
\begin{align}
\mathfrak{C}_{\mathrm{ChSAS}}^{\left(  2n+3\right)  }  &  =\int_{0}%
^{1}\mathrm{d}t\langle nAF_{t}^{n-1}F_{4t}+A_{3}F_{t}^{n}\rangle\nonumber\\
&  =\left\langle F^{n},A_{3}\right\rangle +\mathrm{d}\varphi_{2n+2}.
\label{zehn}%
\end{align}

\subsubsection{ChSAS forms in $(2n+5)$-dimensions}

The third series of exact $(2n+6)$-forms is given by~\cite{savv8}%
\begin{equation}
\Xi_{2n+6}=\langle F^{n},F_{6}\rangle+n\langle F^{n-1},F_{4}^{2}%
\rangle=\mathrm{d}\mathfrak{C}_{\mathrm{ChSAS}}^{\left(  2n+5\right)
},\label{elf}%
\end{equation}
where the corresponding $(2n+5)$-form $\mathfrak{C}_{\mathrm{ChSAS}}^{\left(
2n+5\right)  }$ is defined in terms of the $6$-form $F_{6}=\mathrm{D}%
A_{5}+\left[  A_{3},A_{3}\right]  $ field-strength for the rank-$5$ gauge
field $A_{5}$. As in subsection $2.2.1$ we can now also define the $\left(
2n+5\right)  $-ChSAS form as%
\begin{equation}
\mathfrak{C}_{\mathrm{ChSAS}}^{\left(  2n+5\right)  }=\left\langle F^{n}%
,A_{5}\right\rangle +n\left\langle F^{n-1},F_{4},A_{3}\right\rangle
.\label{zwelf}%
\end{equation}

\subsubsection{ChSAS forms in $(2n+7)$-dimensions}

The fourth series of invariant closed forms $\Gamma_{2n+8}$ in $\left(
2n+8\right)  $ dimensions is given by~\cite{sav5}%
\begin{equation}
\Upsilon_{2n+8}=\langle F^{n},F_{8}\rangle+3n\langle F^{n-1},F_{4}%
,F_{6}\rangle+n(n-1)\langle F^{n-2},F_{4}^{\text{ }3}\rangle=\mathrm{d}%
\mathfrak{C}_{\mathrm{ChSAS}}^{\left(  2n+7\right)  },\label{dreizehn}%
\end{equation}
where the corresponding $(2n+7)$-form $\mathfrak{C}_{\mathrm{ChSAS}}^{\left(
2n+5\right)  }$ is defined in terms of the $8$-form $F_{8}=\mathrm{D}%
A_{7}+3\left[  A_{3},A_{5}\right]  $ field-strength for the rank-$7$ gauge
field $A_{7}$. From \textbf{(}\ref{dreizehn}\textbf{)} it is possible to find
the called $\left(  2n+7\right)  $-ChSAS form%
\begin{equation}
\mathfrak{C}_{\mathrm{ChSAS}}^{\left(  2n+7\right)  }=\left\langle F^{n}%
,A_{7}\right\rangle +n(n-1)\left\langle F_{4},F_{4},A_{3},F^{n-2}\right\rangle
+n\left\langle F_{6},A_{3},F^{n-1}\right\rangle +2n\left\langle F_{4}%
,A_{5},F^{n-1}\right\rangle .\label{vierzehn}%
\end{equation}

\section{Free differential algebras}

In this section, we will make a short review on free differential algebras and
their gauging \cite{sull,dAuria,castell0,castell}.

The dual formulation of Lie algebras provided by the Maurer--Cartan equations
\cite{castell0} can be naturally extended to $p$-forms $\left(  p>1\right)  .$
Let's consider an arbitrary manifold $M$ and a basis of exterior forms
\textbf{\ }$\left\{  \Theta^{A_{1}\left(  p_{1}\right)  },\Theta^{A_{2}\left(
p_{2}\right)  },\ldots,\Theta^{A_{n}\left(  p_{n}\right)  }\right\}  $ defined
on $M$, labeled by the index $A$ and by the degree $p$ of the form, which may
be different for different values of $A$. This means that each $p_{i}$ takes
values $0,1,2,\ldots,N$, while $i$ takes the values $1,2,...,n$.

The external derivative $\mathrm{d}\Theta^{A\left(  p\right)  }$ can be
expressed as a combination of the elements of the base, which leads to write a
generalized Maurer--Cartan equation of the following type
\cite{sull,dAuria,castell0,castell}%
\begin{equation}
\mathrm{d}\Theta^{A\left(  p\right)  }+\sum_{n=1}^{N}\frac{1}{n}%
C_{B_{1}\left(  p_{1}\right)  \cdots B_{n}\left(  p_{n}\right)  }^{A\left(
p\right)  }\Theta^{B_{1}\left(  p_{1}\right)  }\wedge\cdots\wedge\Theta
^{B_{n}\left(  p_{n}\right)  }=0, \label{fda6}%
\end{equation}
where the coefficients $C_{B_{1}\left(  p_{1}\right)  \cdots B_{n}\left(
p_{n}\right)  }^{A\left(  p\right)  }$ are called generalized structure
constants. The symmetry of these constants in the lower index is induced by
the permutation of the forms $\Theta^{A\left(  p\right)  }$ in the product
wedge and are different from zero only if%
\begin{equation}
p_{1}+p_{2}+\cdots+p_{n}=p+1. \label{fda6'}%
\end{equation}
Here, the number $N$ is equal to $p_{\max}+1,$ where $p_{\max}$ is the highest
degree in the set $\left\{  \Theta^{A\left(  p\right)  }\right\}  $. One can
say that Eq. (\ref{fda6}) is a generalized Maurer--Cartan equation and that it
describes a FDA if and only if the integrability condition $\mathrm{d}%
^{2}\Theta^{A\left(  p\right)  }=0$ follows automatically from (\ref{fda6}).
Explicitly, the condition for (\ref{fda6}) to be a FDA is given by%
\begin{align}
\mathrm{d}^{2}\Theta^{A\left(  p\right)  }  &  =\mathrm{-}\sum_{n,m=1}%
^{N}\frac{1}{m}C_{B_{1}\left(  p_{1}\right)  \cdots B_{n}\left(  p_{n}\right)
}^{A\left(  p\right)  }C_{D_{1}\left(  q_{1}\right)  \cdots D_{m}\left(
q_{m}\right)  }^{B_{1}\left(  p_{1}\right)  }\nonumber\\
&  \Theta^{D_{1}\left(  q_{1}\right)  }\wedge\cdots\wedge\Theta^{D_{m}\left(
q_{m}\right)  }\wedge\Theta^{B_{2}\left(  p_{2}\right)  }\wedge\cdots
\wedge\Theta^{B_{n}\left(  p_{n}\right)  }\nonumber\\
&  =0. \label{fda6''}%
\end{align}
This equation is just the analogue of the Jacobi identities of an ordinary Lie
algebra. \ It is very instructive to have a look at the most general form of a
FDA as it emerges from theorems of Sullivan. From Ref. \cite{castell0} we know
that: $\left(  i\right)  $ a FDA is called "minimal algebra" when it is true
that $C_{B(p+1)}^{A(p)}=0$. This means that all forms appearing in the
expansion of $\mathrm{d}\Theta^{A\left(  p\right)  }$ have at most degree $p,$
being the degree $\left(  p+1\right)  $ ruled out; $\left(  ii\right)  $ a FDA
is called a contractible algebra when the only form appearing in the expansion
of $\mathrm{d}\Theta^{A\left(  p\right)  }$ has degree $\left(  p+1\right)  ,$
namely%
\begin{equation}
\mathrm{d}\Theta^{A\left(  p\right)  }=\Theta^{A\left(  p+1\right)  }\text{,
\ i.e., \ }\mathrm{d}\Theta^{A\left(  p+1\right)  }=0.
\end{equation}

\textbf{Sullivan's fundamental theorem: }The most general free differential
algebra is the direct sum of a contractible algebra with a minimal algebra.

\subsection{Gauging free differential algebras}

Physical applications of FDA require a generalization of the concepts of soft
$1$-forms and curvatures introduced gauging Maurer--Cartan equations
\cite{dAuria,castell0,castell}.\textbf{ }

Let $\left\{  A^{B_{1}\left(  p_{1}\right)  },A^{B_{2}\left(  p_{2}\right)
},\ldots,A^{B_{n}\left(  p_{n}\right)  }\right\}  $ be a set of $p$-forms
gauge potential, labeled by the index $B$ and by the degree $p$ of the form,
which may be different for different values of $B$. If we consider the
$p$-forms $A^{B_{i}(p_{i})}$ as the gauge potentials of a FDA, in the same way
as the components $A^{a}$ are the gauge potentials of an ordinary Lie algebra
described by the ordinary Maurer--Cartan equations, then the curvatures
associated with the $A^{B_{i}(p_{i})}$ potentials are given by%
\begin{equation}
F^{A\left(  p+1\right)  }=\mathrm{d}A^{A\left(  p\right)  }+\sum_{n=1}%
^{N}\frac{1}{n}C_{B_{1}\left(  p_{1}\right)  \cdots B_{n}\left(  p_{n}\right)
}^{A\left(  p\right)  }A^{B_{1}\left(  p_{1}\right)  }\wedge\cdots\wedge
A^{B_{n}\left(  p_{n}\right)  }. \label{six}%
\end{equation}
If we apply the exterior derivative to both sides of Eq. (\ref{six}), we
obtain a generalization of the Bianchi identity \cite{castell0}%
\begin{equation}
\nabla F^{A\left(  p+1\right)  }=\mathrm{d}F^{A\left(  p+1\right)  }%
+\sum_{n=1}^{N}C_{B_{1}\left(  p_{1}\right)  \cdots B_{n}\left(  p_{n}\right)
}^{A\left(  p\right)  }F^{B_{1}\left(  p_{1}+1\right)  }\wedge A^{B_{2}\left(
p_{2}\right)  }\wedge\cdots\wedge A^{B_{n}\left(  p_{n}\right)  }=0.
\label{seven}%
\end{equation}
In complete analogy to what one does in ordinary group theory, we say that the
left side of (\ref{seven}) defines the covariant derivative $\nabla$ of an
adjoint set of $(p+1)$-forms. With this definition, the Bianchi identity
(\ref{seven}) just states that the covariant derivative of the curvature set
$F^{A\left(  p+1\right)  }$ is zero as it happens for ordinary groups.

\section{Extended gauge theory and gauged FDA}

Let us now consider the explicit form of the equations (\ref{six}%
,\ref{seven}). \ In the case of a minimal FDA, the explicit form of equations
\textbf{(}\ref{six},\ref{seven}) for\textbf{ }$p=1,2,3,5,7,9,$ is given in
Appendices A and B respectively. Here we will list, using the nomenclature of
Refs. \cite{epjs,sav4,sav5,savv8}, only the equations we will use later. In
fact, from (\ref{Ap1'}) we can see that, if we restrict ourselves to the case
of an FDA whose structure constants satisfy the condition $C_{B(q)C(r)}%
^{A(q+r-1)}=C_{BC}^{\text{ }A}$ for any $r<q$, where $C_{BC}^{\text{ }A}$
correspond to the structure constants of a Lie algebra \footnotetext[1]{We
will consider this condition in the rest of this paper.}, then the equations
(\ref{Ap1}, \ref{Ap1'}) can be written in the form (see Appendix A)%
\begin{align}
F  &  =\mathrm{d}A+A^{2},\nonumber\\
\text{ }F_{3}  &  =\mathrm{d}A_{2}+\left[  A,A_{2}\right]  ,\nonumber\\
F_{4}  &  =\mathrm{d}A_{3}+\left[  A,A_{3}\right]  ,\nonumber\\
F_{6}  &  =\mathrm{d}A_{5}+\left[  A,A_{5}\right]  +\frac{1}{2}\left[
A_{3},A_{3}\right]  ,\nonumber\\
F_{8}  &  =\mathrm{d}A_{7}+\left[  A,A_{7}\right]  +\left[  A_{3}%
,A_{5}\right]  ,\nonumber\\
F_{10}  &  =\mathrm{d}A_{9}+\left[  A,A_{9}\right]  +\left[  A_{3}%
,A_{7}\right]  +\frac{1}{2}\left[  A_{5},A_{5}\right]  . \label{nueve}%
\end{align}

In the same way, for the equation \textbf{(}\ref{Ap2}\textbf{)} we find (see
Appendix B)%
\begin{align}
\mathrm{D}F  &  =0,\nonumber\\
\mathrm{D}F_{3}+\left[  A_{2},F\right]   &  =0,\nonumber\\
\mathrm{D}F_{4}+\left[  A_{3},F\right]   &  =0,\nonumber
\end{align}%
\begin{align}
\mathrm{D}F_{6}+\left[  A_{3},F_{4}\right]  +\left[  A_{5},F\right]   &
=0,\nonumber\\
\mathrm{D}F_{8}+\left[  A_{3},F_{6}\right]  +\left[  A_{5},F_{4}\right]
+\left[  A_{7},F\right]   &  =0,\nonumber\\
\mathrm{D}F_{10}+\left[  A_{3},F_{8}\right]  +\left[  A_{5},F_{6}\right]
+\left[  A_{7},F_{4}\right]  +\left[  A_{9},F\right]   &  =0. \label{diez}%
\end{align}
It should be noted that equations \textbf{(}\ref{nueve}\textbf{)} and
\textbf{(}\ref{diez}\textbf{)} match those found in Refs.
\cite{epjs,sav4,sav5,savv8}, except for numerical coefficients. However, they
coincide exactly after an appropriate transformation of the gauge fields (see
Appendix F).

\subsection{\textbf{Gauge transformations}}

Let $\left\{  \lambda^{B_{1}\left(  p_{1}\right)  },\ldots,\lambda
^{B_{n-1}\left(  p_{n-1}\right)  }\right\}  $ be a set $\left(  p-1\right)
$-forms gauge parameters and let $\left\{  A^{B_{1}\left(  p_{1}\right)
},\ldots,A^{B_{n}\left(  p_{n}\right)  }\right\}  $ be a set of $p$-forms
gauge potentials labeled by an index $B$ and by the degree $p$. Under a gauge
transformation, the gauge potential transforms as%
\begin{equation}
\delta A^{A\left(  p+1\right)  }=\mathrm{d}\lambda^{A\left(  p\right)  }%
+\sum_{n=1}^{N}C_{B_{1}\left(  p_{1}\right)  B_{2}\left(  p_{2}\right)  \cdots
B_{n}\left(  p_{n}\right)  }^{A\left(  p\right)  }A^{B_{1}\left(
p_{1}\right)  }\wedge\lambda^{B_{2}\left(  p_{2}\right)  }\wedge\cdots
\wedge\lambda^{B_{n}\left(  p_{n}\right)  }. \label{nine}%
\end{equation}
In the case of a minimal FDA, the explicit form of equation \textbf{(}%
\ref{nine}) for $n=2$ and \textbf{ }$p=1,2,3,5,7,9,$ is given in Appendix C.
From (\ref{Ap3'}) we can see that%
\begin{align}
\delta A  &  =\mathrm{D}\lambda,\nonumber\\
\delta A_{2}  &  =\mathrm{D}\lambda_{1}+\left[  A_{2},\lambda\right]
,\nonumber\\
\delta A_{3}  &  =\mathrm{D}\lambda_{2}+\left[  A_{3},\lambda\right]
,\nonumber\\
\delta A_{5}  &  =\mathrm{D}\lambda_{4}+\left[  A_{3},\lambda_{2}\right]
+\left[  A_{5},\lambda\right]  ,\nonumber\\
\delta A_{7}  &  =\mathrm{D}\lambda_{6}+\left[  A_{3},\lambda_{4}\right]
+\left[  A_{5},\lambda_{2}\right]  +\left[  A_{7},\lambda\right]  ,\nonumber\\
\delta A_{9}  &  =\mathrm{D}\lambda_{8}+\left[  A_{3},\lambda_{6}\right]
+\left[  A_{5},\lambda_{4}\right]  +\left[  A_{7},\lambda_{2}\right]  +\left[
A_{9},\lambda\right]  . \label{once}%
\end{align}

\subsection{\textbf{Gauge transformations for curvatures}}

Following the definition of the usual gauge theory, we have%
\begin{equation}
\delta F^{A(p+1)}=\nabla\left(  \delta A^{A(p)}\right)  , \label{inv6}%
\end{equation}
so that%
\begin{align}
\delta F^{A\left(  p+1\right)  }  &  =\nabla\left(  \delta A^{A(p)}\right)
=\mathrm{d}\left(  \delta A^{A(p)}\right) \nonumber\\
&  +\sum_{n=1}^{N}C_{B_{1}\left(  p_{1}\right)  B_{2}\left(  p_{2}\right)
\cdots B_{n}\left(  p_{n}\right)  }^{A\left(  p\right)  }\delta A^{B_{1}%
\left(  p_{1}\right)  }\wedge A^{B_{2}\left(  p_{2}\right)  }\wedge
\cdots\wedge A^{B_{n}\left(  p_{n}\right)  }. \label{inv7}%
\end{align}
In the case of a minimal FDA, the explicit form of equation \textbf{(}%
\ref{inv7}) for\textbf{ }$p=1,\ldots,9$ is given Appendix D. When a FDA has
structure constants that satisfy the condition $C_{B(q)C(r)}^{A(q+r-1)}%
=C_{BC}^{\text{ }A}$, we find that the equations (\ref{Ap4'}) can be written
in the form (see Appendix D)%
\begin{align}
\delta F  &  =\left[  F,\lambda\right]  ,\nonumber\\
\delta F_{4}  &  =\left[  F_{4},\lambda\right]  +\left[  F,\lambda_{2}\right]
,\nonumber\\
\delta F_{6}  &  =\left[  F_{6},\lambda\right]  +\left[  F_{4},\lambda
_{2}\right]  +\left[  F,\lambda_{4}\right]  ,\nonumber\\
\delta F_{8}  &  =\left[  F_{8},\lambda\right]  +\left[  F_{6},\lambda
_{2}\right]  +\left[  F_{4},\lambda_{4}\right]  +\left[  F,\lambda_{6}\right]
,\nonumber\\
\delta F_{10}  &  =\left[  F_{10},\lambda\right]  +\left[  F_{8},\lambda
_{2}\right]  +\left[  F_{6},\lambda_{4}\right]  +\left[  F_{4},\lambda
_{6}\right]  +\left[  F,\lambda_{8}\right]  . \label{doce}%
\end{align}
The equations (\ref{once}\textbf{,}\ref{doce}\textbf{)} match those found in
Refs. \cite{epjs,sav4,sav5,savv8}, after an appropriate redefinition of the
gauge fields (see Appendix F).

\section{Extended invariants}

In this section it is shown that the extended invariants found by Antoniadis
and Savvidy in Refs. \cite{epjs,sav4,sav5,savv8} can be constructed from a
gauged free differential algebra.

\subsection{Chern--Pontryagin invariants}

Let $A=A^{a}T_{a}$ be a $1$-form connection evaluated in the Lie algebra
$\mathfrak{g}$ of the group $G$ and let $F=F^{a}T_{a}=\mathrm{d}A+A^{2}$ be
its corresponding $2$-form curvature. The Chern--Pontryagin topological
invariant in $2n+2$ dimensions is given by \cite{zan}%
\begin{equation}
\mathcal{P}_{2n+2}=\left\langle F\wedge\cdots\wedge F\right\rangle
=g_{a_{1}\cdots a_{n+1}}F^{a_{1}}\wedge\cdots\wedge F^{a_{n+1}},
\end{equation}
where the bracket $\left\langle \cdots\right\rangle $ is a symmetric
multilinear form that represents an appropriately normalized trace over the
algebra defined by%
\begin{equation}
g_{a_{1}\cdots a_{n+1}}=\left\langle T_{a_{1}},\ldots,T_{a_{n+1}}\right\rangle
.
\end{equation}

\subsection{Generalized Chern--Pontryagin invariants}

Let's consider now the generalization of the Chern--Pontryagin topological
invariant to the case where Lie algebra $\mathfrak{g}$ is replaced by a free
differential algebra.\ Let $\left\{  F^{B_{1}\left(  p_{1}\right)  }%
,\ldots,F^{B_{n+1}\left(  p_{n+1}\right)  }\right\}  $ be a set of $p$-forms
field intensities. It is possible to construct topological invariants
analogous to the Chern--Pontryagin invariant as follows%
\begin{align}
\mathcal{\tilde{P}}  &  =\sum_{\left\{  p_{i}\right\}  }\left\langle
F^{(p_{1})}\wedge\cdots\wedge F^{(p_{n+1})}\right\rangle \nonumber\\
&  =\sum_{\left\{  p_{i}\right\}  }g_{B_{1}(p_{1})\cdots B_{n+1}(p_{n+1}%
)}F^{B_{1}(p_{1})}\wedge\cdots\wedge F^{B_{n+1}(p_{n+1})}, \label{inv8}%
\end{align}
where for each order of the form $\mathcal{\tilde{P}}$, the sum runs over all
possible combinations.

\subsubsection{\textbf{Case }$p_{1}+\cdots+p_{n+1}=2n+2$}

If $p_{1}+\cdots+p_{n+1}=2n+2$, the only possible choice is $p_{1}%
=\cdots=p_{n+1}=2$. Then we find%
\begin{align}
\mathcal{\tilde{P}}  &  =g_{B_{1}(p_{1})\cdots B_{n}(p_{n+1})}F^{B_{1}(p_{1}%
)}\wedge\cdots\wedge F^{B_{n+1}(p_{n+1})}\nonumber\\
&  =g_{B_{1}(2)\cdots B_{n+1}(2)}F^{B_{1}(2)}\wedge\cdots\wedge F^{B_{n+1}%
(2)}\nonumber\\
&  =\left\langle F^{(2)}\wedge\cdots\wedge F^{(2)}\right\rangle =\left\langle
\left[  F^{(2)}\right]  ^{n+1}\right\rangle .
\end{align}
Using the nomenclature used in Refs. \cite{epjs,sav4,sav5,savv8} we can write%
\begin{equation}
\mathcal{P}=\left\langle F^{n+1}\right\rangle ,
\end{equation}
which coincides with the usual Chern--Pontryagin invariant $\mathcal{P}%
_{2n+2}$.

\subsubsection{\textbf{Case \ }$p_{1}+\cdots+p_{n+1}=2n+3$}

If $p_{1}+\cdots+p_{n+1}=2n+3$ the only possible choice is $p_{1}=\cdots
=p_{n}=2;$ $p_{n+1}=3$. According to the permutations law, there must exist
$n+1$ terms of the form%
\begin{align}
&  g_{B_{1}(2)\cdots B_{n}(2)B_{n+1}(3)}F^{B_{1}(2)}\wedge\cdots\wedge
F^{B_{n}(2)}\wedge F^{B_{n+1}(3)}\nonumber\\
&  =\left\langle F^{(2)}\wedge\cdots\wedge F^{(2)}\wedge F^{(3)}\right\rangle
=\left\langle \left[  F^{(2)}\right]  ^{n},F^{(3)}\right\rangle ,
\end{align}
so that, the corresponding extended Chern--Pontryagin invariant is given by%
\begin{equation}
\mathcal{\tilde{P}}=\left(  n+1\right)  \left\langle F^{(2)}\wedge\cdots\wedge
F^{(2)}\wedge F^{(3)}\right\rangle =\left(  n+1\right)  \left\langle
F^{(2)n},F^{(3)}\right\rangle .
\end{equation}
Using the nomenclature used in Refs. \cite{epjs,sav4,sav5,savv8} we find%
\begin{equation}
\mathcal{P}_{2n+3}=\left\langle F^{n},F_{3}\right\rangle .
\end{equation}
Since $\mathrm{d}\mathcal{P}_{2n+3}=0$ we have $\mathcal{P}_{2n+3}%
=\mathrm{d}\mathfrak{C}^{\left(  2n+2\right)  }$. \ Following the usual
procedure we have%
\begin{equation}
\mathfrak{C}^{\left(  2n+2\right)  }=\left\langle F^{n},A_{2}\right\rangle
+\mathrm{d}\varphi_{2n+1}.
\end{equation}
These\ results coincide with the extended Chern--Pontryagin $\left(
2n+3\right)  $-dimensional and with the $\left(  2n+2\right)  $-Chern--Simons
forms $\mathfrak{C}_{\mathrm{ChSAS}}^{\left(  2n+2\right)  }$ found by
Antoniadis and Savvidy in Refs. \cite{epjs,sav4,sav5}.

\subsubsection{\textbf{Case \ }$p_{1}+\cdots+p_{n+1}=2n+4$}

According to the permutations law, there must exist $n+1$ terms of the form%
\begin{align}
&  g_{B_{1}(2)\cdots B_{n}(2)B_{n+1}(4)}F^{B_{1}(2)}\cdots\wedge F^{B_{n}%
(2)}\wedge F^{B_{n+1}(4)}\nonumber\\
&  =\left\langle F^{(2)}\wedge\cdots\wedge F^{(2)}\wedge F^{(4)}\right\rangle
=\left\langle \left[  F^{(2)}\right]  ^{n},F^{(4)}\right\rangle ,
\end{align}
so that, the corresponding extended Chern--Pontryagin invariant is given by%
\begin{equation}
\mathcal{\tilde{P}}=\left(  n+1\right)  \left\langle F^{(2)}\wedge\cdots\wedge
F^{(2)}\wedge F^{(4)}\right\rangle =\left(  n+1\right)  \left\langle \left[
F^{(2)}\right]  ^{n},F^{(4)}\right\rangle .
\end{equation}
Using the nomenclature used in Refs. \cite{epjs,sav4,sav5,savv8} we can write
as%
\begin{equation}
\mathcal{P}_{2n+4}=\left\langle F^{n},F_{4}\right\rangle .
\end{equation}
Since $\mathrm{d}\mathcal{P}_{2n+4}=0$ we have $\mathcal{P}_{2n+4}%
=\mathrm{d}\mathfrak{C}^{\left(  2n+3\right)  }$, where%
\begin{equation}
\mathfrak{C}^{\left(  2n+3\right)  }=\left\langle F^{n},A_{3}\right\rangle
+\mathrm{d}\varphi_{2n+2}.
\end{equation}
These\ results coincides with the extended topological invariant and with the
$\left(  2n+3\right)  $-Chern--Simons forms $\mathfrak{C}_{\mathrm{ChSAS}%
}^{\left(  2n+3\right)  }$ found in Refs. \cite{epjs,sav4,sav5}.

\subsubsection{\textbf{Case }$p_{1}+\cdots+p_{n+1}=2n+6$}

In this case we will choice two combinations which will be analyze separately.

\paragraph{\textbf{Term with }$p_{1}=\cdots=p_{n}=2$\textbf{ and }$p_{n+1}%
=6:$}

In this case we have that, according to the permutations law, there must exist
$n+1$ terms of the form%
\begin{align}
&  g_{B_{1}(2)\cdots B_{n}(2)B_{n}(6)}F^{B_{1}(2)}\wedge\cdots\wedge
F^{B_{n}(2)}\wedge F^{B_{n+1}(6)}\nonumber\\
&  =\left\langle F^{(2)}\wedge\cdots\wedge F^{(2)}\wedge F^{(6)}\right\rangle
=\left\langle \left[  F^{(2)}\right]  ^{n},F^{(6)}\right\rangle ,
\end{align}
so that%
\begin{equation}
\mathcal{\tilde{P}}_{1}=\left(  n+1\right)  \left\langle F^{(2)}\wedge
\cdots\wedge F^{(2)}\wedge F^{(6)}\right\rangle =\left(  n+1\right)
\left\langle \left[  F^{(2)}\right]  ^{n},F^{(6)}\right\rangle .
\end{equation}

\paragraph{\textbf{Term with }$p_{1}=\cdots=p_{n-1}=2$\textbf{ and }%
$p_{n}=p_{n+1}=4:$}

In this case we have that, according to the law of permutations, there must
exist $n(n+1)/2$ terms of the form%
\begin{align}
&  g_{B_{1}(2)\cdots B_{n-1}(2)B_{n}(4)B_{n+1}(4)}F^{B_{1}(2)}\wedge
\cdots\wedge F^{B_{n-1}(2)}\wedge F^{B_{n}(4)}\wedge F^{B_{n+1}(4)}\nonumber\\
&  =\left\langle F^{(2)}\wedge\cdots\wedge F^{(2)}\wedge F^{(4)}\wedge
F^{(4)}\right\rangle =\left\langle \left[  F^{(2)}\right]  ^{n-1},\left[
F^{(4)}\right]  ^{2}\right\rangle ,
\end{align}
so that%
\begin{equation}
\mathcal{\tilde{P}}_{2}=\frac{n(n+1)}{2}\left\langle \left[  F^{(2)}\right]
^{n-1},\left[  F^{(4)}\right]  ^{2}\right\rangle .
\end{equation}
This means that the corresponding extended Chern--Pontryagin invariant is
given by%
\begin{equation}
\mathcal{\tilde{P}}=\mathcal{\tilde{P}}_{1}+\mathcal{\tilde{P}}_{2}=\left(
n+1\right)  \left\langle \left[  F^{(2)}\right]  ^{n},F^{(6)}\right\rangle
+\frac{n\left(  n+1\right)  }{2}\left\langle \left[  F^{(2)}\right]
^{n-1},\left[  F^{(4)}\right]  ^{2}\right\rangle ,
\end{equation}
which can be write as%
\begin{equation}
\mathcal{P}=\left\langle \left[  F^{(2)}\right]  ^{n},F^{(6)}\right\rangle
+\frac{n}{2}\left\langle \left[  F^{(2)}\right]  ^{n-1},\left[  F^{(4)}%
\right]  ^{2}\right\rangle . \label{intop1}%
\end{equation}
Using the nomenclature used in Refs. \cite{epjs,sav4,sav5,savv8} we can write%
\begin{equation}
\mathcal{P}_{2n+6}=\left\langle F^{n},F_{6}\right\rangle +\frac{n}%
{2}\left\langle F^{n-1},F_{4}^{\text{ }2}\right\rangle . \label{intop1'}%
\end{equation}
Now let us now prove that the expression (\ref{intop1'}) is, in addition to
being gauge invariant, a closed form. The variation of $\mathcal{P}_{2n+6}$ is
given by%
\begin{align}
\delta\mathcal{P}_{2n+6}  &  =n\left\langle F^{n-1},\delta F,F_{6}%
\right\rangle +\left\langle F^{n},\delta F_{6}\right\rangle \nonumber\\
&  +\frac{n\left(  n-1\right)  }{2}\left\langle F^{n-2},\delta F,F_{4}^{\text{
}2}\right\rangle +\frac{n}{2}\left\langle F^{n-1},F_{4},\delta F_{4}%
\right\rangle . \label{intop1''}%
\end{align}
Introducing (\ref{doce}) into (\ref{intop1''}) we have%
\begin{align}
\delta\mathcal{P}_{2n+6}  &  =\left\langle \left[  F_{4},\lambda_{2}\right]
,F^{n}\right\rangle +\left\langle \left[  F,\lambda_{4}\right]  ,F^{n}%
\right\rangle +n\left\langle \left[  F,\lambda_{2}\right]  ,F_{4}%
,F^{n-1}\right\rangle \nonumber\\
&  =\left\{  \left\langle \left[  F_{4},\lambda_{2}\right]  ,F^{n}%
\right\rangle +n\left\langle \left[  F,\lambda_{2}\right]  ,F_{4}%
,F^{n-1}\right\rangle \right\}  +\left\langle \left[  F,\lambda_{4}\right]
,F^{n}\right\rangle \nonumber\\
&  =0.
\end{align}
Now let us show that (\ref{intop1'}) is a closed form. Taking the exterior
derivative of $\mathcal{P}_{2n+6}$ we have%
\begin{align}
\mathrm{d}\mathcal{P}_{2n+6}  &  =\left\langle \mathrm{D}F_{6},F^{n}%
\right\rangle +n\left\langle F_{6},\mathrm{D}F,F^{n-1}\right\rangle
+n\left\langle \mathrm{D}F_{4},F_{4},F^{n-1}\right\rangle \nonumber\\
&  +\frac{n\left(  n-1\right)  }{2}\left\langle F_{4}^{2},\mathrm{D}%
F,F^{n-2}\right\rangle .
\end{align}
Using (\ref{diez}) we have%
\begin{align}
\mathrm{d}\mathcal{P}_{2n+6}  &  =\left\langle \left[  F_{4},A_{3}\right]
,F^{n}\right\rangle +\left\langle \left[  F,A_{5}\right]  ,F^{n}\right\rangle
+n\left\langle \left[  F,A_{3}\right]  ,F_{4},F^{n-1}\right\rangle \nonumber\\
&  =\left\{  \left\langle \left[  F_{4},A_{3}\right]  ,F^{n}\right\rangle
+n\left\langle \left[  F,A_{3}\right]  ,F_{4},F^{n-1}\right\rangle \right\}
+\left\langle \left[  F,A_{5}\right]  ,F^{n}\right\rangle ,
\end{align}
and using the well known identity \cite{zumino}%
\begin{equation}
\sum_{i=1}^{n}(-1)^{(d_{1}+\cdots+d_{i-1})d_{\Theta}}\left\langle \Lambda
_{1},\ldots,\left[  \Theta,\Lambda_{i}\right]  ,\ldots,\Lambda_{n}%
\right\rangle =0, \label{id1}%
\end{equation}
where each $\Lambda_{i}$ is a $d_{i}$-form and $\Theta$ is an arbitrary
$d_{\Theta}$ -form, we have%
\begin{equation}
\mathrm{d}\mathcal{P}_{2n+6}=0,
\end{equation}
which proves that the form $\mathcal{P}_{2n+6}$ is closed. This means that
$\mathcal{P}_{2n+6}=\mathrm{d}\mathfrak{C}^{2n+5}$ where, following the usual
procedure, we find%
\begin{equation}
\mathfrak{C}^{\left(  2n+5\right)  }=\left\langle F^{n},A_{5}\right\rangle
+\frac{n}{2}\left\langle F_{4},A_{3},F^{n-1}\right\rangle +\mathrm{d}%
\varphi_{2n+4}. \label{chs1}%
\end{equation}

\subsubsection{\textbf{Case }$p_{1}+\cdots+p_{n+1}=2n+8$}

In this case we will choice three combinations which will be analyze separately.

\paragraph{\textbf{Term with }$p_{1}=p_{2}=\cdots=p_{n-1}=2$\textbf{ and
}$p_{n}=8:$}

Now we have according to the law of permutations, $n+1$ terms of the form%
\begin{align}
&  g_{B_{1}(2)\cdots B_{n}(2)B_{n+1}(8)}F^{B_{1}(2)}\wedge\cdots\wedge
F^{B_{n}(2)}\wedge F^{B_{n+1}(8)}\nonumber\\
&  =\left\langle F^{(2)}\wedge\cdots\wedge F^{(2)}\wedge F^{(8)}\right\rangle
=\left\langle \left[  F^{(2)}\right]  ^{n},F^{(8)}\right\rangle ,
\end{align}
so that%
\begin{equation}
\mathcal{\tilde{P}}_{1}=\left(  n+1\right)  \left\langle F^{(2)}\wedge
\cdots\wedge F^{(2)}\wedge F^{(8)}\right\rangle =\left(  n+1\right)
\left\langle \left[  F^{(2)}\right]  ^{n},F^{(8)}\right\rangle .
\end{equation}

\paragraph{\textbf{Term with }$p_{1}=\cdots=p_{n-1}=2$\textbf{, }$p_{n}%
=4$\textbf{ and }$p_{n+1}=6:$}

According to the law of permutations, we have $n(n+1)$ terms of the form%
\begin{align}
&  g_{B_{1}(2)\cdots B_{n-1}(2)B_{n}(4)B_{n+1}(6)}F^{B_{1}(2)}\wedge
\cdots\wedge F^{B_{n-1}(2)}\wedge F^{B_{n}(4)}\wedge F^{B_{n+1}(6)}\nonumber\\
&  =\left\langle F^{(2)}\wedge\cdots\wedge F^{(2)}\wedge F^{(4)}\wedge
F^{(6)}\right\rangle =\left\langle \left[  F^{(2)}\right]  ^{n-1}%
,F^{(4)},F^{(6)}\right\rangle ,
\end{align}
so that%
\begin{equation}
\mathcal{\tilde{P}}_{2}=n(n+1)\left\langle F^{(2)}\wedge\cdots\wedge
F^{(2)}\wedge F^{(4)}\wedge F^{(6)}\right\rangle =n(n+1)\left\langle \left[
F^{(2)}\right]  ^{n-1},F^{(4)},F^{(6)}\right\rangle .
\end{equation}

\paragraph{\textbf{Term with }$p_{1}=\cdots=p_{n-2}=2$\textbf{ and }%
$p_{n-1}=p_{n}=p_{n+1}=4:$}

The permutations law, tells us that there are $\left(  n+1\right)  n(n-1)/3!$
terms of the form%
\begin{align}
&  g_{B_{1}(2)\cdots B_{n-2}(2)B_{n-1}(4)B_{n}(4)B_{n+1}(4)}F^{B_{1}(2)}%
\wedge\cdots\wedge F^{B_{n-2}(2)}\wedge F^{B_{n-1}(4)}\wedge F^{B_{n}%
(4)}\wedge F^{B_{n+1}(4)}\nonumber\\
&  =\left\langle F^{(2)}\wedge\cdots\wedge F^{(2)}\wedge F^{(4)}\wedge
F^{(4)}\wedge F^{(4)}\right\rangle =\left\langle \left[  F^{(2)}\right]
^{n-2},\left[  F^{(4)}\right]  ^{3}\right\rangle ,
\end{align}
so that%
\begin{align}
\mathcal{\tilde{P}}_{3}  &  =\frac{n(n+1)(n-1)}{3!}\left\langle F^{(2)}%
\wedge\cdots\wedge F^{(2)}\wedge F^{(4)}\wedge F^{(4)}\wedge F^{(4)}%
\right\rangle \nonumber\\
&  =\frac{n(n+1)(n-1)}{3!}\left\langle \left[  F^{(2)}\right]  ^{n-2},\left[
F^{(4)}\right]  ^{3}\right\rangle .
\end{align}
This means that the corresponding extended Chern--Pontryagin invariant is
given by%
\begin{align}
\mathcal{\tilde{P}}  &  =\mathcal{\tilde{P}}_{1}+\mathcal{\tilde{P}}%
_{2}+\mathcal{\tilde{P}}_{3}\nonumber\\
&  =\left(  n+1\right)  \left\langle \left[  F^{(2)}\right]  ^{n}%
,F^{(8)}\right\rangle +n(n+1)\left\langle \left[  F^{(2)}\right]
^{n-1},F^{(4)},F^{(6)}\right\rangle \nonumber\\
&  +\frac{n(n+1)(n-1)}{3!}\left\langle \left[  F^{(2)}\right]  ^{n-2},\left[
F^{(4)}\right]  ^{3}\right\rangle ,
\end{align}
which can be written as%
\begin{equation}
\mathcal{P}_{2n+8}=\left\langle \left[  F^{(2)}\right]  ^{n},F^{(8)}%
\right\rangle +n\left\langle \left[  F^{(2)}\right]  ^{n-1},F^{(4)}%
,F^{(6)}\right\rangle +\frac{n(n-1)}{3!}\left\langle \left[  F^{(2)}\right]
^{n-2},\left[  F^{(4)}\right]  ^{3}\right\rangle , \label{intop3}%
\end{equation}
that coincides, except for two numerical coefficients, with the topological
invariant $\Gamma_{2n+8}$ found by Antoniadis and Savvidy in Refs.
\cite{epjs,sav4,sav5}.\ Using the nomenclature from these references, the
equation (\ref{intop3}) takes the form%
\begin{equation}
\mathcal{P}_{2n+8}=\left\langle F_{8},F^{n}\right\rangle +n\left\langle
F_{4},F_{6},F^{n-1}\right\rangle +\frac{n(n-1)}{3!}\left\langle F_{4}%
^{3},F^{n-2}\right\rangle . \label{intop4}%
\end{equation}
Now let us prove that (\ref{intop4}) is gauge invariant. The variation of
$\mathcal{P}_{2n+8}$ is given by%
\begin{align}
\delta\mathcal{P}_{2n+8}  &  =\left\langle \delta F_{8},F^{n}\right\rangle
+n\left\langle F_{8},\delta F,F^{n-1}\right\rangle +n\left\langle \delta
F_{4},F_{6},F^{n-1}\right\rangle \nonumber\\
&  +n\left\langle F_{4},\delta F_{6},F^{n-1}\right\rangle +n\left(
n-1\right)  \left\langle F_{4},F_{6},\delta F,F^{n-2}\right\rangle \nonumber\\
&  +\frac{n(n-1)}{2}\left\langle \delta F_{4},F_{4}^{2},F^{n-2}\right\rangle
+\frac{n(n-1)(n-2)}{3!}\left\langle F_{4}^{\text{ }3},\delta F,F^{n-3}%
\right\rangle . \label{intop4'}%
\end{align}
Introducing (\ref{doce}) into (\ref{intop4'}) we find%
\begin{align}
\delta\mathcal{P}_{2n+8}  &  =-\left\{  \left\langle \left[  F_{6},\lambda
_{2}\right]  ,F^{n}\right\rangle +n\left\langle \left[  F,\lambda_{2}\right]
,F_{6},F^{n-1}\right\rangle \right\} \nonumber\\
&  -\frac{n}{2}\left\{  2\left\langle F^{\left(  4\right)  },\left[
F_{4},\lambda_{2}\right]  ,F^{n-1}\right\rangle +(n-1)\left\langle \left[
F_{4},\lambda_{2}\right]  ,F_{4}^{\text{ }2},F^{n-2}\right\rangle \right\}
\nonumber\\
&  -\left\{  \left\langle \left[  F_{4},\lambda_{4}\right]  ,F^{n}%
\right\rangle +n\left\langle F_{4},\left[  F,\lambda_{4}\right]
,F^{n-1}\right\rangle \right\}  +\left\langle \left[  F,\lambda_{6}\right]
,F^{n}\right\rangle \nonumber\\
&  =0.
\end{align}
Let us now show that (\ref{intop4}) is also a closed form. Taking the exterior
derivative of $\mathcal{P}_{2n+8}$ we have%
\begin{align}
\mathrm{d}\mathcal{P}_{2n+8}  &  =\left\langle \mathrm{D}F_{8},F^{n}%
\right\rangle +n\left\langle F_{8},\mathrm{D}F,F^{n-1}\right\rangle
+n\left\langle \mathrm{D}F_{4},F_{6},F^{n-1}\right\rangle \nonumber\\
&  +n\left\langle F_{4},\mathrm{D}F_{6},F^{n-1}\right\rangle +n\left(
n-1\right)  \left\langle F_{4},F_{6},\mathrm{D}F,F^{n-2}\right\rangle
\nonumber\\
&  +\frac{3n(n-1)}{3!}\left\langle \mathrm{D}F_{4},F_{4}^{2},F^{n-2}%
\right\rangle +\frac{n(n-1)(n-2)}{3!}\left\langle F_{4}^{\text{ }3}%
,\mathrm{D}F,F^{n-3}\right\rangle .
\end{align}
Using (\ref{diez}) we find%
\begin{align}
\mathrm{d}\mathcal{P}_{2n+8}  &  =\left\langle \left[  F_{6},A_{3}\right]
,F^{n}\right\rangle +\left\langle \left[  F_{4},A_{5}\right]  ,F^{n}%
\right\rangle +\left\langle \left[  F,A_{7}\right]  ,F^{n}\right\rangle
\nonumber\\
&  +n\left\langle \left[  F,A_{3}\right]  ,F_{6},F^{n-1}\right\rangle
+n\left\langle F_{4},\left[  F_{4},A_{3}\right]  F^{n-1}\right\rangle
\nonumber\\
&  +n\left\langle F_{4},\left[  F,A_{5}\right]  ,F^{n-1}\right\rangle
+\frac{n(n-1)}{2}\left\langle \left[  F_{4},A_{3}\right]  ,F_{4}^{2}%
,F^{n-2}\right\rangle ,
\end{align}
or equivalently%
\begin{align}
\mathrm{d}\mathcal{P}_{2n+8}  &  =\left\{  \left\langle \left[  F_{6}%
,A_{3}\right]  ,F^{n}\right\rangle +n\left\langle \left[  F,A_{3}\right]
,F_{6},F^{n-1}\right\rangle \right\} \nonumber\\
&  +\frac{n}{2}\left\{  2\left\langle F_{4},\left[  F_{4},A_{3}\right]
,F^{n-1}\right\rangle +(n-1)\left\langle \left[  F_{4},A_{3}\right]
,F_{4}^{2},F^{n-2}\right\rangle \right\} \nonumber\\
&  +\left\{  \left\langle \left[  F_{4},A_{5}\right]  ,F^{n}\right\rangle
+n\left\langle F_{4},\left[  F,A_{5}\right]  ,F^{n-1}\right\rangle \right\}
+\left\langle \left[  F,A_{7}\right]  ,F^{n}\right\rangle .
\end{align}
Then using (\ref{id1}) we can see%
\begin{equation}
\mathrm{d}\mathcal{P}_{2n+8}=0,
\end{equation}
which proves that the form $\mathcal{P}_{2n+8}$ is a closed form. This means
that $\mathcal{P}_{2n+8}=\mathrm{d}\mathfrak{C}^{2n+7}$, where $\mathfrak{C}%
^{2n+7}$ is given by%
\begin{align}
\mathfrak{C}^{\left(  2n+7\right)  }  &  =\left\langle F^{n}A_{7}\right\rangle
+\frac{n(n-1)}{3!}\left\langle F_{4},F_{4},A_{3},F^{n-2}\right\rangle
+\frac{n}{3}\left\langle F_{6},A_{3},F^{n-1}\right\rangle \nonumber\\
&  +\frac{2n}{3}\left\langle F_{4},A_{5},F^{n-1}\right\rangle +\mathrm{d}%
\varphi_{2n+6}. \label{chs2}%
\end{align}

The equations (\ref{intop1'},\ref{intop4},\ref{chs1},\ref{chs2}) match those
found in Refs. \cite{epjs,sav4,sav5,savv8}, after an appropriate redefinition
of the gauge fields (see Appendix F).

\section{Concluding Remarks}

In this article we have shown that the so-called ChSAS invariants
\cite{epjs,sav4,sav5,savv8} can be constructed from a algebraic structure
known as gauged free differential algebras.\ The series of exact
$(2n+p)$-forms is given by%
\begin{align*}
\mathcal{P}_{2n+3}  &  =\mathrm{d}\mathfrak{C}^{\left(  2n+2\right)  }=\langle
F^{n},F_{3}\rangle,\\
\mathcal{P}_{2n+4}  &  =\mathrm{d}\mathfrak{C}^{\left(  2n+3\right)  }=\langle
F^{n},F_{4}\rangle,\\
\mathcal{P}_{2n+6}  &  =\mathrm{d}\mathfrak{C}^{\left(  2n+5\right)  }=\langle
F^{n},F_{6}\rangle+\frac{n}{2}\langle F^{n-1},F_{4}^{2}\rangle,\\
\mathcal{P}_{2n+8}  &  =\mathrm{d}\mathfrak{C}^{\left(  2n+7\right)  }=\langle
F^{n},F_{8}\rangle+n\langle F^{n-1},F_{4},F_{6}\rangle+\frac{n(n-1)}%
{3!}\langle F^{n-2},F_{4}^{3}\rangle,
\end{align*}
where each $F_{q+1}$ is a $\left(  q+1\right)  $-form field-strength for the
rank-$q$ gauge field $A_{q}$ which depends also on other gauge fields $A_{r}$
with $r<q.$ The corresponding secondary $(2n+p)$-form $\mathfrak{C}^{\left(
2n+p\right)  }$ are also defined in terms of such gauge fields in the
following way%
\begin{align*}
\mathfrak{C}^{\left(  2n+2\right)  }  &  =\left\langle F^{n},A_{2}%
\right\rangle +\mathrm{d}\varphi_{2n+1},\\
\mathfrak{C}^{\left(  2n+3\right)  }  &  =\left\langle F^{n},A_{3}%
\right\rangle +\mathrm{d}\varphi_{2n+2},\\
\mathfrak{C}^{\left(  2n+5\right)  }  &  =\left\langle F^{n},A_{5}%
\right\rangle +\frac{n}{2}\left\langle F_{4},A_{3},F^{n-1}\right\rangle
+\mathrm{d}\varphi_{2n+4},\\
\mathfrak{C}^{\left(  2n+7\right)  }  &  =\left\langle F^{n},A_{7}%
\right\rangle +\frac{n(n-1)}{3!}\left\langle F_{4},F_{4},A_{3},F^{n-2}%
\right\rangle +\frac{n}{3}\left\langle F_{6},A_{3},F^{n-1}\right\rangle
+\frac{2n}{3}\left\langle F_{4},A_{5},F^{n-1}\right\rangle +\mathrm{d}%
\varphi_{2n+6}.
\end{align*}
If we consider the $n=2$ case in the definition of $\mathfrak{C}^{\left(
2n+7\right)  }$, we find that the $11$-dimensional ChSAS form is given by%
\begin{equation}
\mathfrak{C}^{11}=\left\langle F^{2},A_{7}\right\rangle +\frac{1}%
{3}\left\langle F_{4},F_{4},A_{3}\right\rangle +\frac{2}{3}\left\langle
F_{6},A_{3},F\right\rangle +\frac{4}{3}\left\langle F_{4},A_{5},F\right\rangle
.
\end{equation}
From here we can see that the second term has the same form as a term that
appears in the CJS supergravity \cite{cjs}, whose action is given by\
\begin{align}
S_{11}  &  =\int_{M_{11}}L_{11}\nonumber\\
&  =\int_{M_{11}}-\frac{1}{4}R^{ab}\Sigma_{ab}+\frac{i}{2}\bar{\psi}%
\Gamma_{\left(  8\right)  }\mathrm{D}\psi+\frac{i}{8}\left(  T^{a}-\frac{i}%
{4}\bar{\psi}\Gamma^{a}\psi\right)  e_{a}\bar{\psi}\Gamma_{\left(  6\right)
}\psi\nonumber\\
&  -\frac{1}{2}F_{4}\ast F_{4}+\left(  \ast F_{4}+b\right)  \left(
F-a\right)  +\frac{1}{2}ab-\frac{1}{3}A_{3}F_{4}F_{4},
\end{align}
where%
\begin{align}
\Sigma_{a_{1}\cdots a_{r}}  &  :=\frac{1}{\left(  D-r\right)  !}%
\varepsilon_{a_{1}\cdots a_{r}a_{r+1}\cdots a_{D}}e^{a_{r+1}}\cdots e^{a_{D}%
},\\
\Gamma_{\left(  n\right)  }  &  :=\frac{1}{n!}\Gamma_{a_{1}\cdots a_{n}%
}e^{a_{1}}\cdots e^{a_{n}},\text{ \ }a:=\frac{i}{4}\bar{\psi}\Gamma_{\left(
2\right)  }\psi,\\
b  &  :=\frac{i}{4}\bar{\psi}\Gamma_{\left(  5\right)  }\psi,\text{ \ }%
F_{4}=\mathrm{d}A_{4},
\end{align}
and the $\ast$ symbol denotes the Hodge operator. In fact, if one sets the
metric and gravitino field to zero, 11 dimensional supergravity \cite{cjs} is
reduced to a Chern--Simons like theory based on a three form $A$ whose action
is%
\begin{equation}
S=\int_{M_{11}}A_{3}F_{4}F_{4},
\end{equation}
where $F_{4}$ is a $4$-form and $M_{11}$ is an eleven dimensional manifold.
This result allows us to conjecture that it would be possible to construct a
theory of 11-dimensional Chern--Simons supergravity using a procedure similar
to that shown in Ref. \cite{ims}, which contains or ends at some limit in
standard $11$-dimensional supergravity theory \cite{cjs}.

\begin{acknowledgments}
The authors would like to thank Jos\'{e} M. Izquierdo for stimulating
discussions and for his hospitality at Universidad de Valladolid, where part
of this work was done. This work was supported in part by \textrm{FONDECYT}
grants 1130653 and 1150719 from the Government of Chile. One of the authors
(\textrm{SS}) was supported by grant 21140490 from \textrm{CONICYT}, the
bilateral \textrm{DAAD-CONICYT} grant 62160015 and Universidad de
Concepci\'{o}n, Chile.
\end{acknowledgments}

\section*{Appendix A}

Consider now the explicit form of the equation (\ref{six}) for $n=2$ and
$p=1,2,3,5,7,9$%
\begin{align}
F^{A\left(  2\right)  }  &  =\mathrm{d}A^{A\left(  1\right)  }+\frac{1}%
{2}C_{B_{1}\left(  1\right)  B_{2}\left(  1\right)  }^{A\left(  1\right)
}A^{B_{1}\left(  1\right)  }A^{B_{2}\left(  1\right)  },\nonumber\\
\text{ }F^{A\left(  3\right)  }  &  =\mathrm{d}A^{A\left(  2\right)
}+C_{B_{1}\left(  1\right)  B_{2}\left(  2\right)  }^{A\left(  2\right)
}A^{B_{1}\left(  1\right)  }A^{B_{2}\left(  2\right)  },\nonumber\\
F^{A\left(  4\right)  }  &  =\mathrm{d}A^{A\left(  3\right)  }+C_{B_{1}\left(
1\right)  B_{2}\left(  3\right)  }^{A\left(  3\right)  }A^{B_{1}\left(
1\right)  }A^{B_{2}\left(  3\right)  },\nonumber\\
F^{A\left(  6\right)  }  &  =\mathrm{d}A^{A\left(  5\right)  }+C_{B_{1}\left(
1\right)  B_{2}\left(  5\right)  }^{A\left(  5\right)  }A^{B_{1}\left(
1\right)  }A^{B_{2}\left(  5\right)  }+\frac{1}{2}C_{B_{1}\left(  3\right)
B_{2}\left(  3\right)  }^{A\left(  5\right)  }A^{B_{1}\left(  3\right)
}A^{B_{2}\left(  3\right)  },\nonumber\\
F^{A\left(  8\right)  }  &  =\mathrm{d}A^{A\left(  7\right)  }+C_{B_{1}\left(
1\right)  B_{2}\left(  7\right)  }^{A\left(  7\right)  }A^{B_{1}\left(
1\right)  }A^{B_{2}\left(  7\right)  }+C_{B_{1}\left(  3\right)  B_{2}\left(
5\right)  }^{A\left(  7\right)  }A^{B_{1}\left(  3\right)  }A^{B_{2}\left(
5\right)  },\nonumber\\
F^{A\left(  10\right)  }  &  =\mathrm{d}A^{A\left(  9\right)  }+C_{B_{1}%
\left(  1\right)  B_{2}\left(  9\right)  }^{A\left(  9\right)  }%
A^{B_{1}\left(  1\right)  }A^{B_{2}\left(  9\right)  }+C_{B_{1}\left(
3\right)  B_{2}\left(  7\right)  }^{A\left(  9\right)  }A^{B_{1}\left(
3\right)  }A^{B_{2}\left(  7\right)  }\nonumber\\
&  +\frac{1}{2}C_{B_{1}\left(  5\right)  B_{2}\left(  5\right)  }^{A\left(
9\right)  }A^{B_{1}\left(  5\right)  }A^{B_{2}\left(  5\right)  }, \label{Ap1}%
\end{align}
where from $p=3$ we have considered only odd-order gauge fields. \ Note that
we have considered a FDA whose structure constants satisfy the condition
$C_{B(q)C(r)}^{A(q+r-1)}$ for any $r<q.$\ These equations can be written as%
\begin{align}
F_{2}^{A}  &  =\mathrm{d}A_{1}^{A}+\frac{1}{2}C_{BC}^{A}A_{1}^{B}A_{1}%
^{C},\nonumber\\
F_{3}^{A}  &  =\mathrm{d}A_{2}^{A}+C_{BC}^{A}A_{1}^{B}A_{2}^{C}\text{,}%
\nonumber\\
F_{4}^{A}  &  =\mathrm{d}A_{3}^{A}+C_{BC}^{A}A_{1}^{B}A_{3}^{C},\nonumber\\
F_{6}^{A}  &  =\mathrm{d}A_{5}^{A}+C_{BC}^{A}A_{1}^{B}A_{5}^{C}+\frac{1}%
{2}C_{BC}^{A}A_{3}^{B}A_{3}^{C},\nonumber\\
F_{8}^{A}  &  =\mathrm{d}A_{7}^{A}+C_{BC}^{A}A_{1}^{B}A_{7}^{C}+C_{BC}%
^{A}A_{3}^{B}A_{5}^{C},\nonumber\\
F_{10}^{A}  &  =\mathrm{d}A_{9}^{A}+C_{BC}^{A}A_{1}^{B}A_{9}^{C}+C_{BC}%
^{A}A_{3}^{B}A_{7}^{C}+\frac{1}{2}C_{BC}^{A}A_{5}^{B}A_{5}^{C}. \label{Ap1'}%
\end{align}

\section*{Appendix B}

The explicit form of the equation (\ref{seven}) for $n=2$ and $p=1,2,3,5,7,9$
\ is%
\begin{align*}
\mathrm{d}F^{A\left(  2\right)  }  &  =-C_{B_{1}\left(  1\right)  B_{2}\left(
1\right)  }^{A\left(  1\right)  }F^{B_{1}\left(  2\right)  }A^{B_{2}\left(
1\right)  },\\
\mathrm{d}F^{A\left(  3\right)  }  &  =-C_{B_{1}(1)B_{2}\left(  2\right)
}^{A\left(  2\right)  }F^{B_{1}\left(  2\right)  }A^{B_{2}\left(  2\right)
}-C_{B_{1}\left(  2\right)  B_{2}\left(  1\right)  }^{A\left(  2\right)
}F^{B_{1}\left(  3\right)  }A^{B_{2}\left(  1\right)  },\\
\mathrm{d}F^{A\left(  4\right)  }  &  =-C_{B_{1}\left(  1\right)  B_{2}\left(
3\right)  }^{A\left(  3\right)  }F^{B_{1}\left(  2\right)  }A^{B_{2}\left(
3\right)  }-C_{B_{1}\left(  3\right)  B_{2}\left(  1\right)  }^{A\left(
3\right)  }F^{B_{1}\left(  4\right)  }A^{B_{2}\left(  1\right)  },\\
\mathrm{d}F^{A\left(  6\right)  }  &  =-C_{B_{1}\left(  1\right)  B_{2}\left(
5\right)  }^{A\left(  5\right)  }F^{B_{1}\left(  2\right)  }A^{B_{2}\left(
5\right)  }-C_{B_{1}\left(  3\right)  B_{2}\left(  3\right)  }^{A\left(
5\right)  }F^{B_{1}\left(  4\right)  }A^{B_{2}\left(  3\right)  }\\
&  -C_{B_{1}\left(  5\right)  B_{2}\left(  1\right)  }^{A\left(  5\right)
}F^{B_{1}\left(  6\right)  }A^{B_{2}\left(  1\right)  },
\end{align*}%
\begin{align}
\mathrm{d}F^{A\left(  8\right)  }  &  =-C_{B_{1}\left(  1\right)  B_{2}\left(
7\right)  }^{A\left(  7\right)  }F^{B_{1}\left(  2\right)  }A^{B_{2}\left(
7\right)  }-C_{B_{1}\left(  3\right)  B_{2}\left(  5\right)  }^{A\left(
7\right)  }F^{B_{1}\left(  4\right)  }A^{B_{2}\left(  5\right)  }%
-C_{B_{1}\left(  5\right)  B_{2}\left(  3\right)  }^{A\left(  7\right)
}F^{B_{1}\left(  6\right)  }A^{B_{2}\left(  3\right)  }\nonumber\\
&  -C_{B_{1}\left(  7\right)  B_{2}\left(  1\right)  }^{A\left(  7\right)
}F^{B_{1}\left(  8\right)  }A^{B_{2}\left(  1\right)  },\nonumber\\
\mathrm{d}F^{A\left(  10\right)  }  &  =-C_{B_{1}\left(  1\right)
B_{2}\left(  9\right)  }^{A\left(  9\right)  }F^{B_{1}\left(  2\right)
}A^{B_{2}\left(  9\right)  }-C_{B_{1}\left(  3\right)  B_{2}\left(  7\right)
}^{A\left(  9\right)  }F^{B_{1}\left(  4\right)  }A^{B_{2}\left(  7\right)
}-C_{B_{1}\left(  5\right)  B_{2}\left(  5\right)  }^{A\left(  9\right)
}F^{B_{1}\left(  6\right)  }A^{B_{2}\left(  5\right)  }\nonumber\\
&  -C_{B_{1}\left(  7\right)  B_{2}\left(  3\right)  }^{A\left(  9\right)
}F^{B_{1}\left(  8\right)  }A^{B_{2}\left(  3\right)  }-C_{B_{1}\left(
9\right)  B_{2}\left(  1\right)  }^{A\left(  9\right)  }F^{B_{1}\left(
10\right)  }A^{B_{2}\left(  1\right)  }, \label{Ap2}%
\end{align}
where from $p=3$ we have considered only odd-order gauge fields. \ These
equations can be written as%
\begin{align}
\mathrm{d}F_{2}^{A}  &  =-C_{BC}^{A}A_{1}^{B}F_{2}^{C},\nonumber\\
\mathrm{d}F_{3}^{A}  &  =-C_{BC}^{A}A_{2}^{B}F_{2}^{C}-C_{BC}^{A}A_{1}%
^{B}F_{3}^{C},\nonumber\\
\mathrm{d}F_{4}^{A}  &  =-C_{BC}^{A}A_{3}^{B}F_{2}^{C}-C_{BC}^{A}A_{1}%
^{B}F_{4}^{C},\nonumber\\
\mathrm{d}F_{6}^{A}  &  =-C_{BC}^{A}A_{5}^{B}F_{2}^{C}-C_{BC}^{A}A_{3}%
^{B}F_{4}^{C}-C_{BC}^{A}A_{1}^{B}F_{6}^{C},\nonumber\\
\mathrm{d}F_{8}^{A}  &  =-C_{BC}^{A}A_{7}^{B}F_{2}^{C}-C_{BC}^{A}A_{5}%
^{B}F_{4}^{C}-C_{BC}^{A}A_{3}^{B}F_{6}^{C}-C_{BC}^{A}A_{1}^{B}F_{8}%
^{C},\nonumber\\
\mathrm{d}F_{10}^{A}  &  =-C_{BC}^{A}A_{9}^{B}F_{2}^{C}-C_{BC}^{A}A_{7}%
^{B}F_{4}^{C}-C_{BC}^{A}A_{5}^{B}F_{6}^{C}-C_{BC}^{A}A_{3}^{B}F_{8}^{C}%
-C_{BC}^{A}A_{1}^{B}F_{10}^{C}. \label{Ap2'}%
\end{align}

\section*{Appendix C}

The explicit form of the equation (\ref{nine}) for $n=2$ and $p=1,2,3,5,7,9$
\ is given by%
\begin{align}
\delta A^{A\left(  1\right)  }  &  =\mathrm{d}\lambda^{A\left(  0\right)
}+C_{B_{1}\left(  1\right)  B_{2}\left(  0\right)  }^{A\left(  0\right)
}A^{B_{1}\left(  1\right)  }\lambda^{B_{2}\left(  0\right)  },\nonumber\\
\delta A^{A\left(  2\right)  }  &  =\mathrm{d}\lambda^{A\left(  1\right)
}+C_{B_{1}\left(  1\right)  B_{2}\left(  1\right)  }^{A\left(  1\right)
}A^{B_{1}\left(  1\right)  }\lambda^{B_{2}\left(  1\right)  }+C_{B_{1}\left(
2\right)  B_{2}\left(  0\right)  }^{A\left(  1\right)  }A^{B_{1}\left(
2\right)  }\lambda^{B_{2}\left(  0\right)  },\nonumber\\
\delta A^{A\left(  3\right)  }  &  =\mathrm{d}\lambda^{A\left(  2\right)
}+C_{B_{1}(1)B_{2}\left(  2\right)  }^{A\left(  2\right)  }A^{B_{1}\left(
1\right)  }\lambda^{B_{2}\left(  2\right)  }+C_{B_{1}\left(  3\right)
B_{2}\left(  0\right)  }^{A\left(  2\right)  }A^{B_{1}\left(  3\right)
}\lambda^{B_{2}\left(  0\right)  },\nonumber\\
\delta A^{A\left(  5\right)  }  &  =\mathrm{d}\lambda^{A\left(  4\right)
}+C_{B_{1}\left(  1\right)  B_{2}\left(  4\right)  }^{A\left(  4\right)
}A^{B_{1}\left(  1\right)  }\lambda^{B_{2}\left(  4\right)  }+C_{B_{1}\left(
3\right)  B_{2}\left(  2\right)  }^{A\left(  4\right)  }A^{B_{1}\left(
3\right)  }\lambda^{B_{2}\left(  2\right)  }+C_{B_{1}\left(  5\right)
B_{2}\left(  0\right)  }^{A\left(  4\right)  }A^{B_{1}\left(  5\right)
}\lambda^{B_{2}\left(  0\right)  },\nonumber\\
\delta A^{A\left(  7\right)  }  &  =\mathrm{d}\lambda^{A\left(  6\right)
}+C_{B_{1}\left(  1\right)  B_{2}\left(  6\right)  }^{A\left(  6\right)
}A^{B_{1}\left(  1\right)  }\lambda^{B_{2}\left(  76\right)  }+C_{B_{1}\left(
3\right)  B_{2}\left(  4\right)  }^{A\left(  6\right)  }A^{B_{1}\left(
3\right)  }\lambda^{B_{2}\left(  4\right)  }+C_{B_{1}\left(  5\right)
B_{2}\left(  2\right)  }^{A\left(  6\right)  }A^{B_{1}\left(  5\right)
}\lambda^{B_{2}\left(  2\right)  }\nonumber\\
&  +C_{B_{1}\left(  7\right)  B_{2}\left(  0\right)  }^{A\left(  6\right)
}A^{B_{1}\left(  7\right)  }\lambda^{B_{2}\left(  0\right)  },\nonumber\\
\delta A^{A\left(  9\right)  }  &  =\mathrm{d}\lambda^{A\left(  8\right)
}+C_{B_{1}\left(  1\right)  B_{2}\left(  8\right)  }^{A\left(  8\right)
}A^{B_{1}\left(  1\right)  }\lambda^{B_{2}(8)}+C_{B_{1}\left(  3\right)
B_{2}\left(  6\right)  }^{A\left(  8\right)  }A^{B_{1}\left(  3\right)
}\lambda^{B_{2}\left(  6\right)  }+C_{B_{1}\left(  5\right)  B_{2}\left(
4\right)  }^{A\left(  8\right)  }A^{B_{1}\left(  5\right)  }\lambda
^{B_{2}\left(  4\right)  }\nonumber\\
&  +C_{B_{1}\left(  7\right)  B_{2}\left(  2\right)  }^{A\left(  8\right)
}A^{B_{1}\left(  7\right)  }\lambda^{B_{2}\left(  2\right)  }+C_{B_{1}\left(
9\right)  B_{2}\left(  0\right)  }^{A\left(  8\right)  }A^{B_{1}\left(
9\right)  }\lambda^{B_{2}\left(  0\right)  }, \label{Ap3}%
\end{align}
where from $p=3$ we have considered only odd-order gauge fields. These
equations can be written as%
\begin{align}
\delta A_{1}^{A}  &  =\mathrm{d}\lambda_{0}^{A}+C_{BC}^{A}A_{1}^{B}\lambda
_{0}^{C},\nonumber\\
\delta A_{2}^{A}  &  =\mathrm{d}\lambda_{1}^{A}+C_{BC}^{A}A_{1}^{B}\lambda
_{1}^{C}+C_{BC}^{A}A_{2}^{B}\lambda_{0}^{C},\nonumber\\
\delta A_{3}^{A}  &  =\mathrm{d}\lambda_{2}^{A}+C_{BC}^{A}A_{1}^{B}\lambda
_{2}^{C}+C_{BC}^{A}A_{3}^{B}\lambda_{0}^{C},\nonumber\\
\delta A_{5}^{A}  &  =\mathrm{d}\lambda_{4}^{A}+C_{BC}^{A}A_{1}^{B}\lambda
_{4}^{C}+C_{BC}^{A}A_{3}^{B}\lambda_{2}^{C}+C_{BC}^{A}A_{5}^{B}\lambda_{0}%
^{C},\nonumber\\
\delta A_{7}^{A}  &  =\mathrm{d}\lambda_{6}^{A}+C_{BC}^{A}A_{1}^{B}\lambda
_{6}^{C}+C_{BC}^{A}A_{3}^{B}\lambda_{4}^{C}+C_{BC}^{A}A_{5}^{B}\lambda_{2}%
^{C}+C_{BC}^{A}A_{7}^{B}\lambda_{0}^{C},\nonumber\\
\delta A_{9}^{A}  &  =\mathrm{d}\lambda_{8}^{A}+C_{BC}^{A}A_{1}^{B}\lambda
_{8}^{C}+C_{BC}^{A}A_{3}^{B}\lambda_{6}^{C}+C_{BC}^{A}A_{5}^{B}\lambda_{4}%
^{C}+C_{BC}^{A}A_{7}^{B}\lambda_{2}^{C}+C_{BC}^{A}A_{9}^{B}\lambda_{0}^{C}.
\label{Ap3'}%
\end{align}

\section*{Appendix D}

The explicit form of the equation \textbf{(}\ref{inv7}) for $n=2$ and
$p=1,2,3,5,7,9$ \ is given by%
\begin{align}
\delta F^{A\left(  2\right)  }  &  =\nabla\left(  \delta A^{A(1)}\right)
=\mathrm{d}\left(  \delta A^{A(1)}\right)  +C_{B_{1}\left(  1\right)
B_{2}\left(  1\right)  }^{A\left(  1\right)  }\delta A^{B_{1}\left(  1\right)
}A^{B_{2}\left(  1\right)  },\nonumber\\
\delta F^{A\left(  4\right)  }  &  =\nabla\left(  \delta A^{A(3)}\right)
=\mathrm{d}\left(  \delta A^{A(3)}\right)  +C_{B_{1}\left(  1\right)
B_{2}\left(  3\right)  }^{A\left(  3\right)  }\delta A^{B_{1}\left(  1\right)
}A^{B_{2}\left(  3\right)  }+C_{B_{1}\left(  3\right)  B_{2}\left(  1\right)
}^{A\left(  3\right)  }\delta A^{B_{1}\left(  3\right)  }A^{B_{2}\left(
1\right)  },\nonumber\\
\delta F^{A\left(  6\right)  }  &  =\nabla\left(  \delta A^{A(5)}\right)
=\mathrm{d}\left(  \delta A^{A(5)}\right)  +C_{B_{1}\left(  1\right)
B_{2}\left(  5\right)  }^{A\left(  5\right)  }\delta A^{B_{1}\left(  1\right)
}A^{B_{2}\left(  5\right)  }+C_{B_{1}\left(  3\right)  B_{2}\left(  3\right)
}^{A\left(  5\right)  }\delta A^{B_{1}\left(  3\right)  }A^{B_{2}\left(
3\right)  }\nonumber\\
&  +C_{B_{1}\left(  5\right)  B_{2}\left(  1\right)  }^{A\left(  5\right)
}\delta A^{B_{1}\left(  5\right)  }A^{B_{2}\left(  1\right)  },\nonumber\\
\delta F^{A\left(  8\right)  }  &  =\nabla\left(  \delta A^{A(7)}\right)
=\mathrm{d}\left(  \delta A^{A(7)}\right)  +C_{B_{1}\left(  1\right)
B_{2}\left(  7\right)  }^{A\left(  7\right)  }\delta A^{B_{1}\left(  1\right)
}A^{B_{2}\left(  7\right)  }+C_{B_{1}\left(  3\right)  B_{2}\left(  5\right)
}^{A\left(  7\right)  }\delta A^{B_{1}\left(  3\right)  }A^{B_{2}\left(
5\right)  }\nonumber\\
&  +C_{B_{1}\left(  5\right)  B_{2}\left(  3\right)  }^{A\left(  7\right)
}\delta A^{B_{1}\left(  5\right)  }A^{B_{2}\left(  3\right)  }+C_{B_{1}\left(
7\right)  B_{2}\left(  1\right)  }^{A\left(  7\right)  }\delta A^{B_{1}\left(
7\right)  }A^{B_{2}\left(  1\right)  },\nonumber\\
\delta F^{A\left(  10\right)  }  &  =\nabla\left(  \delta A^{A(9)}\right)
=\mathrm{d}\left(  \delta A^{A(9)}\right)  +C_{B_{1}\left(  1\right)
B_{2}\left(  9\right)  }^{A\left(  9\right)  }\delta A^{B_{1}\left(  1\right)
}A^{B_{2}\left(  9\right)  }+C_{B_{1}\left(  3\right)  B_{2}\left(  7\right)
}^{A\left(  9\right)  }\delta A^{B_{1}\left(  3\right)  }A^{B_{2}\left(
7\right)  }\nonumber\\
&  +C_{B_{1}\left(  5\right)  B_{2}\left(  5\right)  }^{A\left(  9\right)
}\delta A^{B_{1}\left(  5\right)  }A^{B_{2}\left(  5\right)  }+C_{B_{1}\left(
7\right)  B_{2}\left(  3\right)  }^{A\left(  9\right)  }\delta A^{B_{1}\left(
7\right)  }A^{B_{2}\left(  3\right)  }+C_{B_{1}\left(  9\right)  B_{2}\left(
1\right)  }^{A\left(  9\right)  }\delta A^{B_{1}\left(  9\right)  }%
A^{B_{2}\left(  1\right)  }, \label{Ap4}%
\end{align}
where, from $p=3$ we have considered only odd-order gauge fields. These
equations can be written as%
\begin{align}
\delta F_{2}^{A}  &  =\mathrm{d}\left(  \delta A_{1}^{A}\right)  +C_{BC}%
^{A}A_{1}^{B}\delta A_{1}^{C},\nonumber\\
\delta F_{4}^{A}  &  =\mathrm{d}\left(  \delta A_{3}^{A}\right)  +C_{BC}%
^{A}A_{3}^{B}\delta A_{1}^{C}+C_{BC}^{A}A_{1}^{B}\delta A_{3}^{C},\nonumber\\
\delta F_{6}^{A}  &  =\mathrm{d}\left(  \delta A_{5}^{A}\right)  +C_{BC}%
^{A}A_{5}^{B}\delta A_{1}^{C}+C_{BC}^{A}A_{3}^{B}\delta A_{3}^{C}+C_{BC}%
^{A}A_{1}^{B}\delta A_{5}^{C},\nonumber
\end{align}%
\begin{align}
\delta F_{8}^{A}  &  =\mathrm{d}\left(  \delta A_{7}^{A}\right)  +C_{BC}%
^{A}A_{7}^{B}\delta A_{1}^{C}+C_{BC}^{A}A_{5}^{B}\delta A_{3}^{C}+C_{BC}%
^{A}A_{3}^{B}\delta A_{5}^{C}+C_{BC}^{A}A_{1}^{B}\delta A_{7}^{C},\nonumber\\
\delta F_{10}^{A}  &  =\mathrm{d}\left(  \delta A_{9}^{A}\right)  +C_{BC}%
^{A}A_{9}^{B}\delta A_{1}^{C}+C_{BC}^{A}A_{7}^{B}\delta A_{3}^{C}+C_{BC}%
^{A}A_{5}^{B}\delta A_{5}^{C}+C_{BC}^{A}A_{3}^{B}\delta A_{7}^{C}\nonumber\\
&  +C_{BC}^{A}A_{1}^{B}\delta A_{9}^{C}. \label{Ap4'}%
\end{align}

\section*{\textbf{Appendix E}}

In this appendix we show that (\ref{Ap4'}) correspond to homogeneous transformations.

\subsection{Gauge transformation of the rank-2 field strength tensor $F_{2}$}

\ Introducing the first equations of (\ref{Ap3'}) in the first equation of
(\ref{Ap4'}) we have%
\begin{align}
\delta F_{2}^{A}  &  =C_{BC}^{A}\mathrm{d}A_{1}^{B}\lambda_{0}^{C}-C_{BC}%
^{A}A_{1}^{B}\mathrm{d}\lambda_{0}^{C}+C_{BC}^{A}A_{1}^{B}\mathrm{d}%
\lambda_{0}^{C}+C_{BC}^{A}A_{1}^{B}C_{EF}^{C}A_{1}^{E}\lambda_{0}%
^{F}\nonumber\\
&  =C_{BC}^{A}\mathrm{d}A_{1}^{B}\lambda_{0}^{C}+C_{BC}^{A}A_{1}^{B}C_{EF}%
^{C}A_{1}^{E}\lambda_{0}^{F}.
\end{align}
Using the nomenclature of Refs. \cite{epjs,sav4,sav5,savv8}, this equation
takes the form%
\begin{equation}
\delta F_{2}=\left[  \mathrm{d}A_{1}+A_{1}A_{1},\lambda_{0}\right]  =\left[
F_{2},\lambda_{0}\right]  . \label{e3}%
\end{equation}

\subsection{Gauge transformation of the rank-4 field strength tensor $F_{4}$}

\ Introducing the first and third equations of (\ref{Ap3'}) in the second
equation of (\ref{Ap4'}), we have%
\begin{align}
\delta F_{4}^{A}  &  =C_{BC}^{A}\mathrm{d}A_{1}^{B}\lambda_{2}^{C}-C_{BC}%
^{A}A_{1}^{B}\mathrm{d}\lambda_{2}^{C}+C_{BC}^{A}\mathrm{d}A_{3}^{B}%
\lambda_{0}^{C}-C_{BC}^{A}A_{3}^{B}\mathrm{d}\lambda_{0}^{C}\nonumber\\
&  +C_{BC}^{A}A_{3}^{B}d\lambda_{0}^{C}+C_{BC}^{A}A_{3}^{B}C_{EF}^{C}A_{1}%
^{E}\lambda_{0}^{F}+C_{BC}^{A}A_{1}^{B}d\lambda_{2}^{C}\nonumber\\
&  +C_{BC}^{A}A_{1}^{B}C_{EF}^{C}A_{1}^{E}\lambda_{2}^{F}+C_{BC}^{A}A_{1}%
^{B}C_{EF}^{C}A_{3}^{E}\lambda_{0}^{F}\nonumber\\
&  =C_{BC}^{A}\mathrm{d}A_{1}^{B}\lambda_{2}^{C}+C_{BC}^{A}\mathrm{d}A_{3}%
^{B}\lambda_{0}^{C}+C_{BC}^{A}A_{3}^{B}C_{EF}^{C}A_{1}^{E}\lambda_{0}%
^{F}\nonumber\\
&  +C_{BC}^{A}A_{1}^{B}C_{EF}^{C}A_{1}^{E}\lambda_{2}^{F}+C_{BC}^{A}A_{1}%
^{B}C_{EF}^{C}A_{3}^{E}\lambda_{0}^{F}. \label{e4}%
\end{align}
Using the nomenclature of Refs. \cite{epjs,sav4,sav5,savv8}, this equation
takes the form%
\begin{align}
\delta F_{4}  &  =\left[  \mathrm{d}A_{1},\lambda_{2}\right]  +\left[
\mathrm{d}A_{3},\lambda_{0}\right]  +\left[  A_{3},\left[  A_{1},\lambda
_{0}\right]  \right]  +\left[  A_{1},\left[  A_{1},\lambda_{2}\right]
\right]  +\left[  A_{1},\left[  A_{3},\lambda_{0}\right]  \right] \nonumber\\
&  =\left[  \mathrm{d}A_{1}+A_{1}A_{1},\lambda_{2}\right]  +\left[
\mathrm{d}A_{3}+\left\{  A_{1}A_{3}\right\}  ,\lambda_{0}\right] \nonumber\\
&  =\left[  F_{2},\lambda_{2}\right]  +\left[  F_{4},\lambda_{0}\right]  .
\label{e5}%
\end{align}

\subsection{Gauge transformation of the rank-4 field strength tensor $F_{6}$}

\ Introducing the first, third and fifth equations of (\ref{Ap3'}) in the
third equation of (\ref{Ap4'}), we have%
\begin{align*}
\delta F_{6}^{A}  &  =C_{BC}^{A}\mathrm{d}A_{1}^{B}\lambda_{4}^{C}-C_{BC}%
^{A}A_{1}^{B}\mathrm{d}\lambda_{4}^{C}+C_{BC}^{A}\mathrm{d}A_{3}^{B}%
\lambda_{2}^{C}-C_{BC}^{A}A_{3}^{B}\mathrm{d}\lambda_{2}^{C}+C_{BC}%
^{A}\mathrm{d}A_{5}^{B}\lambda_{0}^{C}-C_{BC}^{A}A_{5}^{B}\mathrm{d}%
\lambda_{0}^{C}\\
&  +C_{BC}^{A}A_{5}^{B}d\lambda_{0}^{C}+C_{BC}^{A}A_{5}^{B}C_{EF}^{C}A_{1}%
^{E}\lambda_{0}^{F}+C_{BC}^{A}A_{3}^{B}d\lambda_{2}^{C}+C_{BC}^{A}A_{3}%
^{B}C_{EF}^{C}A_{1}^{E}\lambda_{2}^{F}+C_{BC}^{A}A_{3}^{B}C_{EF}^{C}A_{3}%
^{E}\lambda_{0}^{F}\\
&  +C_{BC}^{A}A_{1}^{B}d\lambda_{4}^{C}+C_{BC}^{A}A_{1}^{B}C_{EF}^{C}A_{1}%
^{E}\lambda_{4}^{F}+C_{BC}^{A}A_{1}^{B}C_{EF}^{C}A_{3}^{E}\lambda_{2}%
^{F}+C_{BC}^{A}A_{1}^{B}C_{EF}^{C}A_{5}^{E}\lambda_{0}^{F}\\
&  =C_{BC}^{A}\mathrm{d}A_{1}^{B}\lambda_{4}^{C}+C_{BC}^{A}\mathrm{d}A_{3}%
^{B}\lambda_{2}^{C}+C_{BC}^{A}\mathrm{d}A_{5}^{B}\lambda_{0}^{C}+C_{BC}%
^{A}A_{5}^{B}C_{EF}^{C}A_{1}^{E}\lambda_{0}^{F}+C_{BC}^{A}A_{3}^{B}C_{EF}%
^{C}A_{1}^{E}\lambda_{2}^{F}\\
&  +C_{BC}^{A}A_{3}^{B}C_{EF}^{C}A_{3}^{E}\lambda_{0}^{F}+C_{BC}^{A}A_{1}%
^{B}C_{EF}^{C}A_{1}^{E}\lambda_{4}^{F}+C_{BC}^{A}A_{1}^{B}C_{EF}^{C}A_{3}%
^{E}\lambda_{2}^{F}+C_{BC}^{A}A_{1}^{B}C_{EF}^{C}A_{5}^{E}\lambda_{0}^{F}.
\end{align*}
Using the nomenclature of Refs. \cite{epjs,sav4,sav5,savv8} this equation
takes the form%
\begin{align*}
\delta F_{6}  &  =\left[  \mathrm{d}A_{1},\lambda_{4}\right]  +\left[
\mathrm{d}A_{3},\lambda_{2}\right]  +\left[  \mathrm{d}A_{5},\lambda
_{0}\right]  +\left[  A_{5},\left[  A_{1},\lambda_{0}\right]  \right]
+\left[  A_{3},\left[  A_{1},\lambda_{2}\right]  \right] \\
&  +\left[  A_{3},\left[  A_{3},\lambda_{0}\right]  \right]  +\left[
A_{1},\left[  A_{1},\lambda_{4}\right]  \right]  +\left[  A_{1},\left[
A_{3},\lambda_{2}\right]  \right]  +\left[  A_{1},\left[  A_{5},\lambda
_{0}\right]  \right]  ,
\end{align*}
so that%
\begin{align*}
\delta F_{6}  &  =\left[  \mathrm{d}A_{1}+A_{1}A_{1},\lambda_{4}\right]
+\left[  \mathrm{d}A_{3}+\left[  A_{1},A_{3}\right]  ,\lambda_{2}\right]
+\left[  \mathrm{d}A_{5}+\left[  A_{1},A_{5}\right]  +\frac{1}{2}\left[
A_{3},A_{3}\right]  ,\lambda_{0}\right] \\
&  =\left[  F_{6},\lambda_{0}\right]  +\left[  F_{4},\lambda_{2}\right]
+\left[  F_{2},\lambda_{4}\right]  .
\end{align*}

\subsection{\textbf{Gauge transformation of the field strength tensor }$F_{8}%
$}

Introducing the first, third, fifth and seventh equations of (\ref{Ap3'}) in
the fourth equation of (\ref{Ap4'}), we have%
\begin{align*}
\delta F_{8}^{A}  &  =C_{BC}^{A}\mathrm{d}A_{1}^{B}\lambda_{6}^{C}+C_{BC}%
^{A}\mathrm{d}A_{3}^{B}\lambda_{4}^{C}+C_{BC}^{A}\mathrm{d}A_{5}^{B}%
\lambda_{2}^{C}+C_{BC}^{A}\mathrm{d}A_{7}^{B}\lambda_{0}^{C}+C_{BC}^{A}%
A_{7}^{B}C_{EF}^{C}A_{1}^{E}\lambda_{0}^{F}\\
&  +C_{BC}^{A}A_{5}^{B}C_{EF}^{C}A_{1}^{E}\lambda_{2}^{F}+C_{BC}^{A}A_{5}%
^{B}C_{EF}^{C}A_{3}^{E}\lambda_{0}^{F}+C_{BC}^{A}A_{3}^{B}C_{EF}^{C}A_{1}%
^{E}\lambda_{4}^{F}+C_{BC}^{A}A_{3}^{B}C_{EF}^{C}A_{3}^{E}\lambda_{2}^{F}\\
&  +C_{BC}^{A}A_{3}^{B}C_{EF}^{C}A_{5}^{E}\lambda_{0}^{F}+C_{BC}^{A}A_{1}%
^{B}C_{EF}^{C}A_{1}^{E}\lambda_{6}^{F}+C_{BC}^{A}A_{1}^{B}C_{EF}^{C}A_{3}%
^{E}\lambda_{4}^{F}+C_{BC}^{A}A_{1}^{B}C_{EF}^{C}A_{5}^{E}\lambda_{2}^{F}\\
&  +C_{BC}^{A}A_{1}^{B}C_{EF}^{C}A_{7}^{E}\lambda_{0}^{F}.
\end{align*}
Using the nomenclature of Refs. $\left[  \text{1-4}\right]  $, this equation
takes the form%
\begin{align*}
\delta F_{8}  &  =\left[  \mathrm{d}A_{1},\lambda_{6}\right]  +\left[
\mathrm{d}A_{3},\lambda_{4}\right]  +\left[  \mathrm{d}A_{5},\lambda
_{2}\right]  +\left[  \mathrm{d}A_{7},\lambda_{0}\right]  +\left[
A_{7},\left[  A_{1},\lambda_{0}\right]  \right]  +\left[  A_{5},\left[
A_{1},\lambda_{2}\right]  \right] \\
&  +\left[  A_{5},\left[  A_{3},\lambda_{0}\right]  \right]  +\left[
A_{3},\left[  A_{1},\lambda_{4}\right]  \right]  +\left[  A_{3},\left[
A_{3},\lambda_{2}\right]  \right]  +\left[  A_{3},\left[  A_{5},\lambda
_{0}\right]  \right]  +\left[  A_{1},\left[  A_{1},\lambda_{6}\right]  \right]
\\
&  +\left[  A_{1},\left[  A_{3},\lambda_{4}\right]  \right]  +\left[
A_{1},\left[  A_{5},\lambda_{2}\right]  \right]  +\left[  A_{1},\left[
A_{7},\lambda_{0}\right]  \right]  ,
\end{align*}
so that%
\begin{align*}
\delta F_{8}  &  =\left[  \mathrm{d}A_{1}+A_{1}A_{1},\lambda_{6}\right]
+\left[  \mathrm{d}A_{3}+\left[  A_{1},A_{3}\right]  ,\lambda_{4}\right]
+\left[  \mathrm{d}A_{5}+\left[  A_{1},A_{5}\right]  +\frac{1}{2}\left[
A_{3},A_{3}\right]  ,\lambda_{2}\right] \\
&  +\left[  \mathrm{d}A_{7}+\left[  A_{3},A_{5}\right]  +\left[  A_{1}%
,A_{7}\right]  ,\lambda_{0}\right]  =\left[  F_{8},\lambda\right]  +\left[
F_{6},\lambda_{2}\right]  +\left[  F_{4},\lambda_{4}\right]  +\left[
F,\lambda_{6}\right]  .
\end{align*}

\subsection{\textbf{Gauge transformation of the field strength tensor }%
$F_{10}$}

Introducing the first, third, fifth, seventh and ninth equations from
(\ref{Ap3'}) in the fifth equation of (\ref{Ap4'}) we have%
\begin{align*}
\delta F_{10}^{A}  &  =C_{BC}^{A}\mathrm{d}A_{1}^{B}\lambda_{8}^{C}+C_{BC}%
^{A}\mathrm{d}A_{3}^{B}\lambda_{6}^{C}+C_{BC}^{A}\mathrm{d}A_{5}^{B}%
\lambda_{4}^{C}+C_{BC}^{A}\mathrm{d}A_{7}^{B}\lambda_{2}^{C}+C_{BC}%
^{A}\mathrm{d}A_{9}^{B}\lambda_{0}^{C}\\
&  +C_{BC}^{A}A_{9}^{B}C_{EF}^{C}A_{1}^{E}\lambda_{0}^{F}+C_{BC}^{A}A_{7}%
^{B}C_{EF}^{C}A_{1}^{E}\lambda_{2}^{F}+C_{BC}^{A}A_{7}^{B}C_{EF}^{C}A_{3}%
^{E}\lambda_{0}^{F}+C_{BC}^{A}A_{5}^{B}C_{EF}^{C}A_{1}^{E}\lambda_{4}^{F}\\
&  +C_{BC}^{A}A_{5}^{B}C_{EF}^{C}A_{3}^{E}\lambda_{2}^{F}+C_{BC}^{A}A_{5}%
^{B}C_{EF}^{C}A_{5}^{E}\lambda_{0}^{F}+C_{BC}^{A}A_{3}^{B}C_{EF}^{C}A_{1}%
^{E}\lambda_{6}^{F}+C_{BC}^{A}A_{3}^{B}C_{EF}^{C}A_{3}^{E}\lambda_{4}^{F}\\
&  +C_{BC}^{A}A_{3}^{B}C_{EF}^{C}A_{5}^{E}\lambda_{2}^{F}+C_{BC}^{A}A_{3}%
^{B}C_{EF}^{C}A_{7}^{E}\lambda_{0}^{F}+C_{BC}^{A}A_{1}^{B}C_{EF}^{C}A_{1}%
^{E}\lambda_{8}^{F}+C_{BC}^{A}A_{1}^{B}C_{EF}^{C}A_{3}^{E}\lambda_{6}^{F}\\
&  +C_{BC}^{A}A_{1}^{B}C_{EF}^{C}A_{5}^{E}\lambda_{4}^{F}+C_{BC}^{A}A_{1}%
^{B}C_{EF}^{C}A_{7}^{E}\lambda_{2}^{F}+C_{BC}^{A}A_{1}^{B}C_{EF}^{C}A_{9}%
^{E}\lambda_{0}^{F}.
\end{align*}

Using the nomenclature of Refs. $\left[  \text{1-4}\right]  $, this equation
takes the form%
\begin{align*}
\delta F_{10}  &  =\left[  \mathrm{d}A_{1},\lambda_{8}\right]  +\left[
\mathrm{d}A_{3},\lambda_{6}\right]  +\left[  \mathrm{d}A_{5},\lambda
_{4}\right]  +\left[  \mathrm{d}A_{7},\lambda_{2}\right]  +\left[
\mathrm{d}A_{9},\lambda_{0}\right]  +\left[  A_{9},\left[  A_{1},\lambda
_{0}\right]  \right] \\
&  +\left[  A_{7},\left[  A_{1},\lambda_{2}\right]  \right]  +\left[
A_{7},\left[  A_{3},\lambda_{0}\right]  \right]  +\left[  A_{5},\left[
A_{1},\lambda_{4}\right]  \right]  +\left[  A_{5},\left[  A_{3},\lambda
_{2}\right]  \right]  +\left[  A_{5},\left[  A_{5},\lambda_{0}\right]  \right]
\\
&  +\left[  A_{3},\left[  A_{1},\lambda_{6}\right]  \right]  +\left[
A_{3},\left[  A_{3},\lambda_{4}\right]  \right]  +\left[  A_{3},\left[
A_{5},\lambda_{2}\right]  \right]  +\left[  A_{3},\left[  A_{7},\lambda
_{0}\right]  \right]  +\left[  A_{1},\left[  A_{1},\lambda_{8}\right]  \right]
\\
&  +\left[  A_{1},\left[  A_{3},\lambda_{6}\right]  \right]  +\left[
A_{1},\left[  A_{5},\lambda_{4}\right]  \right]  +\left[  A_{1},\left[
A_{7},\lambda_{2}\right]  \right]  +\left[  A_{1},\left[  A_{9},\lambda
_{0}\right]  \right]  ,
\end{align*}

so that%
\begin{align*}
\delta F_{10}  &  =\left[  \mathrm{d}A_{1}+A_{1}A_{1},\lambda_{8}\right]
+\left[  \mathrm{d}A_{3}+\left[  A_{1},A_{3}\right]  ,\lambda_{6}\right]
+\left[  \mathrm{d}A_{5}+\left[  A_{1},A_{5}\right]  +\frac{1}{2}\left[
A_{3},A_{3}\right]  ,\lambda_{4}\right] \\
&  +\left[  \mathrm{d}A_{7}+\left[  A_{3},A_{5}\right]  +\left[  A_{1}%
,A_{7}\right]  ,\lambda_{2}\right]  +\left[  \mathrm{d}A_{9}+\left[
A_{1},A_{9}\right]  +\left[  A_{3},A_{7}\right]  +\frac{1}{2}\left[
A_{5},A_{5}\right]  ,\lambda_{0}\right] \\
&  =\left[  F_{10},\lambda_{0}\right]  +\left[  F_{8},\lambda_{2}\right]
+\left[  F_{6},\lambda_{4}\right]  +\left[  F_{4},\lambda_{6}\right]  +\left[
F_{2},\lambda_{8}\right]  .
\end{align*}

\section*{\textbf{Appendix F}}

It is interesting to note that the difference between the coefficients that
accompany the terms of equations (\ref{nueve}, \ref{diez}, \ref{once},
\ref{doce}) of this article and the coefficients of the corresponding
equations of Refs. \cite{epjs,sav4,sav5,savv8} can be understood as follows.
Consider the FDA given by Eq. (\ref{fda6}), which leads to the definition of
curvature given by Eq. (\ref{six}). This last equation was restricted to the
case where the only nonzero structure constants are those with only two low
indices. This means that equations (\ref{fda6}) and (\ref{six}) take the form%
\[
\mathrm{d}\Theta^{A\left(  p\right)  }+\frac{1}{2}C_{B_{1}\left(
p_{1}\right)  B_{2}\left(  p_{2}\right)  }^{A\left(  p\right)  }\Theta
^{B_{1}\left(  p_{1}\right)  }\wedge\Theta^{B_{2}\left(  p_{2}\right)  }=0,
\]%
\begin{equation}
F^{A\left(  p+1\right)  }=\mathrm{d}A^{A\left(  p\right)  }+\frac{1}%
{2}C_{B_{1}\left(  p_{1}\right)  B_{2}\left(  p_{2}\right)  }^{A\left(
p\right)  }A^{B_{1}\left(  p_{1}\right)  }\wedge A^{B_{2}\left(  p_{2}\right)
}. \label{efes}%
\end{equation}
The next step is to consider that all the structure constants of the FDA
(\ref{efes}) can be written in terms of the structure constants $C_{B_{1}%
B_{2}}^{A}$ of a Lie algebra. This allows us to write the Eq. (\ref{efes}) in
the form shown in equations (\ref{Ap2}) and (\ref{nueve}).

We have seen that: (i) the generalized field strength tensors transform
homogeneously and (ii) the generalized Chern-Pontryagin invariants are
polynomials in the fields strength tensors and are invariant under gauged and
diffeomorphism transformations. \ Since this invariance is maintained under a
linear redefinition of the tensor gauge fields it is direct to prove that the
fields strength tensor found in the Eqs. (\ref{Ap2}) and (\ref{nueve}) can be
mapped into the field strength tensors defined in Refs.
\cite{epjs,sav4,sav5,savv8}. \ In fact, defining the extended gauged fields of
the following form%
\begin{align}
A  &  \longrightarrow\bar{A}=A;\text{ }A_{3}\longrightarrow\bar{A}_{3}%
=aA_{3};\text{ }A_{5}\longrightarrow\bar{A}_{5}=2aA_{5},\nonumber\\
\text{ }A_{7}  &  \longrightarrow\bar{A}_{7}=6a^{3}A_{7};\text{ }%
A_{9}\longrightarrow\bar{A}_{9}=24a^{4}A_{9}, \label{tr0}%
\end{align}
where $a$ is an arbitrary number, we found that Eq. (\ref{Ap2}) takes the form%
\begin{align}
\bar{F}_{2}^{A}  &  =\mathrm{d}\bar{A}_{1}^{A}+\frac{1}{2}C_{BC}^{A}\bar
{A}_{1}^{B}\bar{A}_{1}^{C},\nonumber\\
\bar{F}_{3}^{A}  &  =\mathrm{d}\bar{A}_{2}^{A}+\frac{1}{2}C_{BC}^{A}\bar
{A}_{1}^{B}\bar{A}_{2}^{C},\nonumber\\
\bar{F}_{4}^{A}  &  =aF_{4}^{A}=\mathrm{d}A_{3}^{A}+C_{BC}^{A}\bar{A}_{1}%
^{B}\bar{A}_{3}^{C},\nonumber\\
\bar{F}_{6}^{A}  &  =2a^{2}F_{6}^{A}=\mathrm{d}\bar{A}_{5}^{A}+C_{BC}^{A}%
\bar{A}_{1}^{B}\bar{A}_{5}^{C}+C_{BC}^{A}\bar{A}_{3}^{B}\bar{A}_{3}%
^{C},\nonumber\\
\bar{F}_{8}^{A}  &  =6a^{3}F_{8}^{A}=\mathrm{d}\bar{A}_{7}^{A}+C_{BC}^{A}%
\bar{A}_{1}^{B}\bar{A}_{7}^{C}+3C_{BC}^{A}\bar{A}_{3}^{B}\bar{A}_{5}%
^{C},\nonumber\\
\bar{F}_{10}^{A}  &  =24a^{4}F_{10}^{A}=\mathrm{d}\bar{A}_{9}^{A}+C_{BC}%
^{A}\bar{A}_{1}^{B}\bar{A}_{9}^{C}+4C_{BC}^{A}\bar{A}_{3}^{B}\bar{A}_{7}%
^{C}+3C_{BC}^{A}\bar{A}_{5}^{B}\bar{A}_{5}^{C}, \label{tr1}%
\end{align}
which can be written in the form%
\begin{align}
\bar{F}  &  =\mathrm{d}\bar{A}+\bar{A}^{2},\nonumber\\
\text{ }\bar{F}_{3}  &  =\mathrm{d}\bar{A}_{2}+\left[  \bar{A},\bar{A}%
_{2}\right]  ,\nonumber\\
\bar{F}_{4}  &  =\mathrm{d}\bar{A}_{3}+\left[  \bar{A},\bar{A}_{3}\right]
,\nonumber\\
\bar{F}_{6}  &  =\mathrm{d}\bar{A}_{5}+\left[  \bar{A},\bar{A}_{5}\right]
+\left[  \bar{A}_{3},\bar{A}_{3}\right]  ,\nonumber\\
\bar{F}_{8}  &  =\mathrm{d}\bar{A}_{7}+\left[  \bar{A},\bar{A}_{7}\right]
+3\left[  \bar{A}_{3},\bar{A}_{5}\right]  ,\nonumber\\
\bar{F}_{10}  &  =\mathrm{d}\bar{A}_{9}+\left[  \bar{A},\bar{A}_{9}\right]
+4\left[  \bar{A}_{3},\bar{A}_{7}\right]  +3\left[  \bar{A}_{5},\bar{A}%
_{5}\right]  . \label{tr2}%
\end{align}
These equations coincide exactly with the equations $\left(  A2\right)  $ of
Ref. \cite{sav5}.

The equations (A5,A1,A4) of Ref. \cite{sav5} can be obtained in an analogous
way. In fact, taking into account that the transformations (\ref{tr0}) induce
in the field strengths and in the gauge parameters the transformations%
\begin{align}
\bar{F}  &  =F,\text{ \ }\bar{F}_{4}=aF_{4},\text{ \ }\bar{F}_{6}=2a^{2}%
F_{6},\nonumber\\
\bar{F}_{8}  &  =6a^{3}F_{8},\text{ \ }\bar{F}_{10}=24a^{4}F_{10}, \label{tr6}%
\end{align}%
\begin{align}
\bar{\lambda}  &  =\lambda,\text{ \ }\bar{\lambda}_{4}=a\lambda_{4},\text{
\ }\bar{\lambda}_{6}=2a^{2}\lambda_{6},\nonumber\\
\bar{\lambda}_{8}  &  =6a^{3}\lambda_{8},\text{ \ }\bar{\lambda}_{10}%
=24a^{4}\lambda_{10}, \label{tr7'}%
\end{align}
it is straightforward to find that the equations (\ref{diez},\ref{once}%
,\ref{doce}) take the form from Eqs. (A5,A1,A4) of Ref. \cite{sav5}. \ In the
same way we can see that, after using the Eqs. (\ref{tr0},\ref{tr6}%
,\ref{tr7'}), the Eqs. (\ref{intop1'},\ref{intop4},\ref{chs1},\ref{chs2}) take
the form of Eqs. $\left(  1.7,1.8,2.15,3.10\right)  $ in Ref. \cite{sav5}
.


\newpage

\begin{thebibliography}{99}                                                                                               %


\bibitem {epjs}I. Antoniadis and G. Savvidy, Eur. Phys. J. C (2012)
\textbf{72} 2140.

\bibitem {sav4}I. Antoniadis and G. Savvidy, Int. Jour. Mod. Phys. A 29 (2014) 1450027.

\bibitem {sav5}S. Konitopoulos and G. Savvidy, Jour. Math. Phys. 55 (2014) 062304.

\bibitem {savv8}G. Savvidy, In. J. Mod. Phys. A \textbf{29} (2014) 1450027.

\bibitem {savv1}G. Savvidy,\ Int. J. Mod. Phys. A \textbf{21} (2006) 4931.

\bibitem {savv2}G. Savvidy, Phys. Lett. B \textbf{625} (2005) 341.

\bibitem {savv3}S. Konitopoulos and G. Savvidy, \ J. Phys. A \textbf{41}
(2008) 355402.

\bibitem {ims}F. Izaurieta, I. Mu\~{n}oz, P. Salgado, Phys. Lett. B
\textbf{750} (2015) 39.

\bibitem {teit}C. Teitelboim, Phys. Lett. B \textbf{167} (1986) 63.

\bibitem {Nakahara}M.~Nakahara, \textit{Geometry, Topology and Physics}.
Institute of Physics Publishing; 2nd edition (2003).

\bibitem {sull}D. Sullivan, \textit{Infinitesimal computations in topology},
Bull. de L'Institut des Hautes Etudes Scientifiques, Publ. Math. 47 (1977).

\bibitem {dAuria}R. D'Auria and P. Fr\'{e}, Nucl. Phys. B \textbf{201} (1982) 101.

\bibitem {castell0}L. Castellani, R. D'Auria and P. Fr\'{e},
\textit{Supergravity and Superstring: a geometric perspective}, World
Scientific, Singapore 1991.

\bibitem {castell}L. Castellani and A. Perotto, Lett. Math. Phys. 38 (1996) 321.

\bibitem {zan}J. Zanelli, \textit{Lectures notes on Chern--Simons
(super)gravities. Second edition (February) 2008}. arXiv:hep-th/0502193.

\bibitem {zumino}B. Zumino, \textit{Chiral Anomalies and Differential
Geometry}, Lectures given at Les Houches, August 1983.

\bibitem {cjs}E. Cremmer, B. Julia and J. Scherk, Phys. Lett. B \textbf{76}
(1978) 409.
\end{thebibliography}
\end{document}